\documentclass[aps,pre,reprint,superscriptaddress,showpacs,showkeys]{revtex4-1} 
\usepackage[utf8]{inputenc}

\usepackage{cmap} 
\usepackage[T1]{fontenc}
\usepackage{microtype}
\usepackage{amsmath,amssymb,mathtools}
\usepackage{txfonts}

\usepackage{mathrsfs}
\usepackage{mleftright}

\usepackage{graphicx,grffile}

\usepackage{textcomp}
\usepackage{siunitx}
\usepackage{booktabs,tabularx,dcolumn}
\newcolumntype{d}[1]{D{.}{.}{#1}}

\usepackage{url,hyperref}
\usepackage[usenames,dvipsnames]{xcolor}
\hypersetup{colorlinks=true, linkcolor=BrickRed, urlcolor=blue!50!black, citecolor=blue!50!black}

\usepackage[capitalize]{cleveref}
\crefrangelabelformat{equation}{(#3#1#4)$-$(#5#2#6)}

\newcommand{\kB}{k_\text{B}}
\newcommand\xA{x_\text{A}}
\newcommand\xB{x_\text{B}}
\newcommand\Dm{D_\text{m}}

\renewcommand\leq{\leqslant}
\renewcommand\geq{\geqslant}
\renewcommand\rho\varrho
\renewcommand\vec[1]{\textrm{\bfseries #1}}

\newcommand\diff{\mathrm{d}}
\newcommand\expect[1]{\left\langle{#1}\right\rangle}
\newcommand\e{\text{e}}
\renewcommand\i{\text{i}}

\DeclareMathOperator\tr{tr}
\DeclareMathOperator\Real{Re}

\begin{document}
\title{Structure and dynamics of binary liquid mixtures near their continuous demixing transitions}

\author{Sutapa Roy}
\email{sutapa@is.mpg.de}
\affiliation{Max-Planck-Institut f\"{u}r Intelligente Systeme,
   Heisenbergstr.\ 3,
   70569 Stuttgart,
   Germany,
   and \\
   IV. Institut f\"{u}r Theoretische Physik,
   Universität Stuttgart,
   Pfaffenwaldring 57,
   70569 Stuttgart,
   Germany}

\author{S. Dietrich}
\affiliation{Max-Planck-Institut f\"{u}r Intelligente Systeme,
   Heisenbergstr.\ 3,
   70569 Stuttgart,
   Germany,
   and \\
   IV. Institut f\"{u}r Theoretische Physik,
   Universität Stuttgart,
   Pfaffenwaldring 57,
   70569 Stuttgart,
   Germany}

\author{Felix Höf{}ling}
\affiliation{Max-Planck-Institut f\"{u}r Intelligente Systeme,
   Heisenbergstr.\ 3,
   70569 Stuttgart,
   Germany,
   and \\
   IV. Institut f\"{u}r Theoretische Physik,
   Universität Stuttgart,
   Pfaffenwaldring 57,
   70569 Stuttgart,
   Germany}
\affiliation{Fachbereich Mathematik und Informatik, Freie Universität Berlin, Arnimallee 6, 14195 Berlin, Germany}

\date{\today}

\begin{abstract}
The dynamic and static critical behavior of five binary Lennard-Jones liquid mixtures, close to their continuous 
demixing points (belonging to the so-called model~$H'$ dynamic universality class), are studied computationally 
by combining semi-grand canonical Monte Carlo simulations and large-scale molecular dynamics (MD) simulations, 
accelerated by graphic processing units (GPU). The symmetric binary liquid mixtures considered cover a variety of 
densities, a wide range of compressibilities, and various interactions between the unlike particles. The static 
quantities studied here encompass the bulk phase diagram (including both the binodal and the $\lambda$-line), the 
correlation length, the concentration susceptibility, the compressibility of the finite-sized systems at the bulk 
critical temperature $T_c$, and the pressure. Concerning the collective transport properties, we focus on the 
Onsager coefficient and the shear viscosity. The critical power-law singularities of these quantities are analyzed 
in the mixed phase (above $T_c$) and non-universal critical amplitudes are extracted. Two universal amplitude ratios 
are calculated. The first one involves static amplitudes only and agrees well with the expectations for the 
three-dimensional Ising universality class. The second ratio includes also dynamic critical amplitudes and is related 
to the Einstein--Kawasaki relation for the interdiffusion constant. Precise estimates of this amplitude ratio are 
difficult to obtain from MD simulations, but within the error bars our results are compatible with theoretical 
predictions and experimental values for model~$H'$. Evidence is reported for an inverse proportionality of the pressure 
and the isothermal compressibility at the demixing transition, upon varying either the number density or the repulsion 
strength between unlike particles.
\end{abstract}

\maketitle

\section{Introduction}
\label{sec:introduction}

Upon approaching continuous phase transitions at $T_c$, the order parameter fluctuations with long wavelengths become 
prevalent \cite{fisher1967, stanley1971}. This is accompanied by the unlimited increase of the bulk correlation 
length $\xi(\tau\to 0^\pm ) \simeq \xi_0^\pm |\tau|^{-\nu}$, where $\tau = (T-T_c)/T_c$ is the reduced temperature 
and $\nu$ is one of the standard bulk critical exponents. The coefficients $\xi_0^\pm$ are \textit{per se} non-universal 
amplitudes but they form a universal ratio $\xi_0^+/\xi_0^-$. This divergence of $\xi$ leads to singularities in and 
scaling behavior of various thermodynamic and transport properties, commonly known as critical phenomena \cite{fisher1967, 
stanley1971, hohenberg1977,privman1991, pelissetto2002, folk2006}. Close to $T_c$ and in line with renormalization 
group theory \cite{fisher1998} the corresponding critical exponents and scaling functions turn out to be universal, 
i.e., they depend only on gross features such as the spatial dimension $d$, the symmetry of the order parameter, the 
range of interactions, and hydrodynamic conservation laws, forming universality classes. Binary liquid mixtures, 
exhibiting second order demixing transitions, serve as experimentally particularly suitable representatives of the 
corresponding Ising universality class (see, e.g., Refs.~\cite{pelissetto2002,heller1967,sengers2009}). Besides probing 
critical phenomena as such, recently these critical demixing transitions in confined binary liquid mixtures have gained 
significant renewed attention in the context of critical Casimir forces \cite{hertlein2008} and of non-equilibrium active 
Brownian motion of colloidal particles, driven by diffusiophoresis in binary liquid solvents~\cite{buttinoni2012}.

Static critical phenomena in binary liquid mixtures are rather well understood and reported in the literature, 
encompassing theory~\cite{fisher1967, stanley1971, hohenberg1977, privman1991, pelissetto2002}, experiments (see, e.g., 
Refs. \cite{heller1967, anisimov2000, sengers2009}), and computer simulations (see, e.g., Refs. \cite{binder2010, binder1981, 
landau2009, wilding1996}). Comparatively, much less is known about their dynamic properties. In particular, simulation 
studies of dynamic critical phenomena are very recent and scarce. So far, most of the computational studies of critical 
transport properties in binary liquid mixtures have been focused on a specific and single, highly incompressible fluid. 
However, probing the concept of universality and its onset for these kind of systems requires simulations of various 
distinct binary liquid mixtures. In order to alleviate this dearth, we have performed MD simulations concerning the 
universal critical behavior \cite{privman1991, hohenberg1977, folk2006, pelissetto2002} of several static and dynamic 
quantities, for five symmetric binary liquid mixtures. The fluids considered here exhibit distinct number densities $\rho$ 
and compressibilities and cover a broad range of critical temperatures. Inter alia, this allows us to investigate the 
density dependence of certain critical amplitudes which together with critical exponents determine the qualitative 
importance of the corresponding critical singularities. In this respect our study is supposed to shed light on apparently 
contradictory theoretical predictions ~\cite{onuki1997, bhattacharjee2010} concerning critical amplitude ratios, for which 
there is also a lack of experimental data. Besides the interest in them in its own right, they play also an important role 
for the dynamics of critical Casimir forces \cite{furukawa2013}, the understanding of which is still in an early stage. The 
compilation of non-universal critical amplitudes, which has been obtained from our study, will also be beneficial for future 
simulation studies involving the demixing of binary liquids, for which one can then select the most appropriate fluid model. 
The MD simulations are carried out in the mixed phase, i.e., approaching $T_c$ from above (for an upper critical demixing 
point); to the best of our knowledge, there are no computational investigations of the critical transport in binary liquid 
mixtures approaching $T_c$ from below. In addition we have estimated certain universal relations \cite{privman1991, hohenberg1977} 
involving various critical amplitudes above $T_c$.

The crucial feature of the static critical phenomena is the unlimited increase of the aforementioned correlation length $\xi$, 
which is a measure of the spatial extent of a typical order parameter fluctuation. In a binary liquid mixture, for a 
demixing transition the order parameter is the deviation of the local concentration from its critical value whereas 
for a liquid--vapor transition the order parameter is the deviation of the local number density from its critical value. 
The static critical singularities obey power laws:
\begin{equation}\label{statics1}
\begin{gathered}
\varphi (\tau \to 0^-) \simeq \varphi_0 |\tau|^\beta, \quad
\xi \simeq \xi_0^\pm |\tau|^{-\nu}, \\
\chi \simeq \chi_0^\pm |\tau|^{-\gamma}, \quad
C_V \simeq A^\pm |\tau|^{-\alpha},
\end{gathered}
\end{equation}
where $\varphi$, $\chi$, and $C_V$ are the order parameter, the susceptibility, and the specific heat at constant volume, 
respectively. Any two of these static critical exponents are independent; all remaining ones follow from scaling relations 
\cite{stanley1971} such as
\begin{equation}\label{staticscaling}
\alpha+2\beta+\gamma=2, \qquad \nu d =2-\alpha ,
\end{equation}
where $d$ is the spatial dimension. Unlike the critical exponents, the critical amplitudes depend on whether $T_c$ is 
approached from above or from below. For the Ising universality class in $d=3$, the exponents are known to high 
accuracy~\cite{pelissetto2002}:
\begin{equation}\label{staticexponentvalue}
\alpha \approx 0.110, \: \beta \approx 0.325, \: \gamma \approx 1.239, \: \text{and} \: \: \nu \approx 0.630.
\end{equation}
The critical amplitudes, both above and below $T_c$, are non-universal. However, certain ratios of the critical amplitudes, 
such as $\xi_0^+/\xi_0^-$ and $\chi_0^+/\chi_0^-$, are known to be universal~\cite{privman1991, anisimov2000}; for the 
Ising universality class in $d=3$ one has
\begin{equation}
\xi_0^+/\xi_0^- \approx 2.02 \; \text{and} \; \chi_0^+/\chi_0^- \approx 4.9.
\end{equation}

Dynamic critical phenomena are governed by the relaxation time $t_r$ which diverges upon approaching $T_c$ as
\begin{equation}
t_r \sim \xi^z,
\end{equation}
leading to critical slowing down \cite{hohenberg1977}. This entails thermal singularities in various collective transport 
coefficients, e.g., the mutual diffusivity $\Dm$ and the shear viscosity $\bar\eta$
\cite{folk2006}:
\begin{equation}\label{dynamics1}
\Dm \sim \xi^{-x_D} \quad \text{and} \quad \bar\eta \sim \xi^{x_\eta}.
\end{equation}
The dynamic critical exponents $z$, $x_D$, and $x_\eta$ satisfy scaling relations as well~\cite{hohenberg1977, folk2006},
\begin{equation}\label{dynamic_relations}
 x_D = d-2 + x_\eta \quad \text{and} \quad z = d + x_\eta \,,
\end{equation}
leaving scope for only one independent dynamic critical exponent. In the case of a liquid--vapor transition, the quantity 
analogous to $\Dm$ is the thermal conductivity $D_T$, bearing the same critical exponent as $\tau \to 0$.

Transport mechanisms in near-critical fluids have to respect hydrodynamic conservation laws, specifically for mass, momentum, 
and internal energy. The dynamics of one-component fluids undergoing a liquid--vapor transition is described by the so-called 
model~$H$~\cite{folk2006, onuki2002}, which incorporates the conservation of a scalar order parameter and of the transverse 
part of the momentum current. The asymptotic behaviors for $\tau \to 0$ of the thermal conductivity and of the shear 
viscosity in model~$H$ have been well studied and the corresponding critical exponents for this universality class are 
known~\cite{folk2006}. A binary liquid mixture, on the other hand, can exhibit two kinds of transitions: liquid--vapor 
transitions at plait points and demixing transitions at consolute points. Unlike a one-component fluid, for a binary liquid 
mixture there are two conserved scalar fields, viz., the concentration field and the density field. Critical dynamics in 
binary liquid mixtures are described by model~$H'$~\cite{folk2006}. The transport properties are 
reported~\cite{folk1995, *folk1995a} to exhibit different features at the consolute and the plait points. For example, 
for $\tau \to 0$ the thermal conductivity remains finite at consolute points, but diverges at plait points of a binary 
liquid mixture~\cite{folk1995, *folk1995a, filippov1968}. Moreover, also the corresponding behaviors in the non-asymptotic 
regime are quite different. However, in the asymptotic regimes the leading critical exponents (including the dynamic ones) 
at the consolute points of a binary mixture are the same as the ones for a one-component fluid~\cite{folk2006}. The to date 
best estimate for $x_\eta\approx 0.068$ in $d=3$ was obtained within a self-consistent mode-coupling approximation for a 
one-component fluid~\cite{hao2005} and is in agreement with previous theoretical calculations~\cite{privman1991, folk1995, 
*folk1995a, onuki1997, bhattacharjee1982, paladin1982, halperin1976}; it is also corroborated by experiments on xenon near 
its liquid--vapor critical point~\cite{berg1999}. By virtue of universality and using \cref{dynamic_relations}, this implies 
for the dynamic critical exponents of model~$H'$ in $d=3$~\cite{folk2006}:
\begin{equation} \label{dynamic_exponents}
  x_\eta \approx 0.068\,, \: x_D \approx 1.068 \,, \: \text{and} \: z \approx 3.068\,.
\end{equation}
The presence of the density as a secondary fluctuating field, which is coupled to the order parameter field (i.e., the 
concentration) through some constraint [cf. \cref{fluctuating_fields}], generally raises the question whether Fisher 
renormalization \cite{fisher1968} has to be accounted for. For the present study of symmetric binary mixtures, we have no 
indications that this is the case, but this issue deserves further theoretical investigation.

Compared with the large body of research on static critical phenomena, there are relatively few studies on dynamic critical 
phenomena. Concerning theory, they are performed mainly by using \textit{mode-coupling} theory (MCT)~\cite{ohta1975, 
kadanoff1968} or \textit{dynamic renormalization group} theory(RGT)~\cite{wilson1972, bausch1976} (see Refs.~\cite{folk2006, 
bhattacharjee2010} for recent reviews). In parallel to that, the phenomenological \textit{dynamic scaling} formalism 
\cite{burstyn1983} has also been used extensively. There are important experimental observations (see, e.g., Refs. 
\cite{sengers2009, swift1968, gillis2005, berg1999}) which have pushed the development of this research area.

On the other hand, there are only few computational studies on dynamic critical phenomena \cite{jagannathan2004, chen2005, 
roy2011}. Such kind of computer simulations for fluids started only a decade ago. The first MD simulation~\cite{jagannathan2004} 
aiming at the critical singularities in the fluid transport quantities was performed in 2004. Although this study 
produced the correct values of the static critical exponents for the susceptibility and the correlation length, the 
reported critical exponent for the interdiffusivity was in disagreement with theoretical predictions. There are numerical 
studies ~\cite{meier2005, dyer2007} of shear and bulk viscosities close to the liquid--vapor transitions of one-component 
fluids too, but, without characterizing quantitatively their critical singularities. The first quantitative determinations 
of critical exponents and amplitudes for transport in fluids---being in accordance with MCT, dynamic RGT, and experiments---were 
performed by Das \textit{et al.} in 2006 by using MD simulations \cite{das2006prl, das2006jcp, das2007jcp}. Applying 
finite-size scaling theory \cite{fisher1971}, the critical singularities of the shear viscosity, of the Onsager coefficient, 
and of the mutual diffusivity were determined. Along these lines, the critical divergence of the bulk viscosity was 
determined recently~\cite{roy2011, roy2013} for a demixing phase transition. In this context, we are aware of only one 
simulation study~\cite{chen2005} of the dynamic critical exponents associated with liquid--vapor transitions (model~$H$). 
Also, there are no studies of dynamic critical phenomena below $T_c$. The latter ones are complicated by non-standard finite-size 
effects changing the location of the binodals \cite{roy2013}, along which the transport quantities have to be calculated.

Simulations of dynamic critical phenomena face particular challenges such as, inter alia, critical slowing down~\cite{onuki1997} 
and finite-size effects. While upon increasing the system size finite-size effects become less pronounced, the critical slowing 
down ($t_r \sim \xi^{z} \sim \tau^{-\nu z} \to \infty$) causes simulations of large systems to become expensive due to increasing 
equilibration times. This leads to noisy simulation data for any transport property near criticality at which large scale 
fluctuations are unavoidable and thus make the determination of critical singularities very difficult. This problem is even 
more pronounced for quantities associated with collective dynamics, such as the shear viscosity, which lack the self-averaging 
of tagged-particle quantities. Critical slowing down also manifests itself in long-time tails of the Green-Kubo correlators of 
transport quantities. In particular, for the bulk viscosity, this has been demonstrated~\cite{meier2005, roy2011, roy2013} to 
make the computation notoriously difficult. Moreover, there are also technical hurdles concerning the temperature 
control~\cite{frenkel2002, roy2014} during long MD runs near $T_c$. One way of dealing with these problems is to carry out 
MD simulations of smaller systems and then to apply a finite-size scaling analysis, as done in Refs. \cite{das2006prl, 
das2006jcp, roy2011, roy2013}. However, in the present study we deal with huge system sizes such that the use of finite-size 
scaling is less important for determining the relevant critical singularities.

This study is organized such that in \cref{sec:model} various models considered here and the simulation methodologies are 
described. \cref{sec:results} contains the results for various static and dynamic quantities. There, we also compare our 
computational observations with available theoretical and experimental predictions. Finally, in \cref{sec:conclusion}, 
we provide a summary and perspectives.

\section{Models and Methods}
\label{sec:model}

\subsection{Models}

As model fluids, we have considered binary mixtures of $\text{A}$ and $\text{B}$ particles, which interact via the 
Lennard-Jones (LJ) pair potential
\begin{equation}\label{LJ1}
  u^\text{LJ}(r;\varepsilon,\sigma)= 4\varepsilon \left[(\sigma/r)^{12} - (\sigma/r)^{6}\right] .
\end{equation}
Particles of species $\alpha,\beta\in\{ \text{A}, \text{B}\}$ have different interaction strengths 
$\varepsilon_{\alpha\beta}$, while for reasons of simplicity all particles share the same diameter $\sigma$ and mass $m$. 
The actually employed pair potentials are
\begin{equation}\label{LJ2}
  u_{\alpha\beta}(r)=\bigl[u_{\alpha \beta}^\text{LJ}(r;\varepsilon_{\alpha\beta},\sigma)-
  u_{\alpha \beta}^\text{LJ}(r_c;\varepsilon_{\alpha\beta},\sigma)\bigr] f\mleft(\frac{r-r_c}{h}\mright) \,,
\end{equation}
where the potential is smoothly truncated at a suitable cut-off distance $r_c$ for computational benefits such that the 
pair force is still continuously differentiable at $r=r_c$. We used the smoothing function 
$f(x)=x^4 \theta(-x)/\bigl(1+x^4\bigr)$, where $\theta$ is the Heaviside step function~\cite{voigtmann2009,zausch2010}. 
A small value of $h=0.005\sigma$ is sufficient to ensure very good numerical stability with respect to conservation laws 
during long MD runs~\cite{colberg2011}, which is indispensable for the study of critical dynamics of molecular fluids.

Throughout, we have used cubic simulation boxes of edge length $L$ and volume $V=L^3$ with periodic boundary conditions 
applied along all Cartesian directions. The total number density $\rho=N/V$ is kept constant, where $N=N_\textrm{A}+N_\textrm{B}$ 
is the total number of particles and $N_{\alpha}$ is the number of particles of species~$\alpha$. With this, the concentration 
is defined as $x_\alpha=N_\alpha/N$. We adopt $\varepsilon_\textrm{AA}=\varepsilon$ as the unit of energy. In turn this 
sets the dimensionless temperature $T^{*}={\kB T}/{\varepsilon}$. For the choice $\varepsilon_\textrm{AA}=\varepsilon_\textrm{BB}$ 
the binary liquid mixture is symmetric. This symmetry leads to several computational advantages concerning the calculation 
of the phase diagram~\cite{landau2009} and improves the statistics of single-particle averages. The various fluids considered 
here are specified by their set of parameters $(\varepsilon_\text{AB}$, $r_c$, $\rho$) to be described next.

As to model~I, we choose
\begin{equation}\label{model1}
r_{c,\alpha\beta} =2.5\sigma, \qquad
\varepsilon_\textrm{AA}=\varepsilon_\textrm{BB}=\varepsilon, \qquad
\varepsilon_\textrm{AB}=\frac{\varepsilon}{2},
\end{equation}
and study various number densities~$\rho$. In model~II, we set $r_{c,\alpha\alpha}=2.5\sigma$ for like-particle interactions 
($\alpha=\beta$) and $r_c=2^{1/6}\sigma$ otherwise, such that the unlike particles interact via the purely repulsive 
Weeks--Chandler--Andersen (WCA) potential~\cite{weeks1971}. For this model, we fix
\begin{equation}
  \rho\sigma^3 =0.8, \qquad \varepsilon_\textrm{AA} =\varepsilon_\textrm{BB}=\varepsilon \,,
\end{equation}
and keep $\varepsilon_\textrm{AB}/\varepsilon$ as a tunable interaction parameter. Model~II is inspired by the Widom--Rowlinson 
mixture~\cite{widom1970}, the dynamics of which has been the subject of a recent simulation study~\cite{jagannathan2004}. The 
Lennard-Jones potential [\cref{LJ1}], and certainly its truncated form [\cref{LJ2}], decay faster than $r^{-(d+2)}$ as 
$r\to \infty$ for $d=3$, which justifies that the critical singularities of static bulk properties belong to the universality 
class of 3D Ising models with short-ranged interactions.

\subsection{Semi-grand canonical Monte Carlo simulation}

The phase diagram and the static susceptibility $\chi$ are calculated using the semi-grand canonical Monte Carlo (SGMC)~\cite{landau2009} 
simulations. Within SGMC, the total particle number $N$ is kept constant, while $\xA$ and $x_B$ fluctuate. The implementation 
of the simulation consists of two Monte Carlo (MC) moves: particle displacement and an identity switch 
$\text{A} \rightleftharpoons \text{B}$. Due to the identity switch also the chemical potential difference 
$\Delta \mu=\mu_\text{A}- \mu_\text{B}$ for the two species enters into the Boltzmann factor. However, for symmetric binary 
liquid mixtures the coexistence curve below $T_c$ is given by $\Delta \mu =0$. Accordingly, it is natural to collect the 
simulation data above $T_c$ also for $\Delta \mu =0$, which implies that the field conjugate to the demixing order parameter 
is zero. During the SGMC runs, the concentration $\xA$ has been recorded at sufficiently large intervals of $10^4$ MC steps, 
so that subsequent samples are approximately independent of each other. For $L=27$ and $\rho\sigma^3=1.0$, the fluid mixture 
has been equilibrated over $3\times 10^6$ MC steps. The attempted particle displacements have been chosen uniformly from the 
cube $[-\sigma/20, \sigma/20] \times [-\sigma/20, \sigma/20] \times [-\sigma/20, \sigma/20]$.

In the SGMC simulation the concentrations $\xA$ and $x_\text{B}$ fluctuate. Therefore the order parameter field 
$\phi := (\xA-\xB)/2$ is not conserved, but the total number density field is conserved. Accordingly, the SGMC dynamics is 
classified as the so-called model~$C$ with a scalar order parameter~\cite{folk2006} and is associated with a dynamic exponent 
$z_\text{SGMC} = 2+\alpha/\nu \approx 2.175$ (see Table~3 in Ref.~\cite{folk2006}). On the other hand, if one performs MC 
simulations in the canonical ensemble ($N_A$, $N_B$, $V$, and $T$ fixed) with rules such that the order parameter is conserved 
\textit{locally}- in addition to a conserved density- it corresponds to model~$D$. Note that both model D and B correspond to a 
\textit{locally} conserved order parameter field, with an additional non-critical conserved density field present in model D. 
It has been shown that the dynamic exponent $z$ for both model B and D as defined above is the same ($z=4-\eta \approx 3.964$) 
\cite{tauber2014,folk2006}. Due to a much smaller value of $z$, MC simulations for model C are computationally faster and more 
advantageous than for model D. However, the issue of MC simulations in the canonical ensemble with an only globally (not locally) 
conserved order parameter field, in the presence of a conserved density field, requires further investigations.

\subsection{Molecular dynamics simulation}

Transport quantities have been calculated by using MD simulations~\cite{allen1987, frenkel2002}, which solve Newton's equations 
of motion for the fluid particles within the microcanonical ensemble ($N_\text{A}, N_\text{B},V$, total momentum, and total 
energy $E$ fixed). Generically, simulations of critical dynamics are challenged both by critical slowing down and by finite-size 
effects~\cite{onuki1997, landau2009}. Even more so, the study of collective transport requires a sufficient separation of length 
scales ($\sigma \ll \xi \ll L$). Therefore, we have performed simulations of very large system sizes ($L \leqslant 50\sigma$) 
containing up to $N=87{,}500$ particles with the trajectories spanning more than $10^4 t_0$ in time, with 
$t_0=\sqrt{m\sigma^2/\varepsilon}$, which is at the high end of the present state of the art.

Such demanding computations have become feasible only recently based on the highly parallel architecture of so-called GPU 
(graphic processing units) accelerators, which are specialized on streaming numerical computations of large data sets in parallel. 
The success of GPU computing in the realm of MD simulations~\cite{baker2011} has stimulated the development of GPU implementations 
for more advanced algorithms~\cite{weigel2012}, and today such accelerator hardware is often part of new installations in 
high-performance computing centers. Specifically, we have used the software \emph{HAL's MD package} 
(version 1.0)~\cite{colberg2011, HALMD}, which is a high-precision molecular dynamics package for large-scale simulations of 
complex dynamics in inhomogeneous liquids. The implementation achieves excellent conservation of energy and momentum at high 
performance by using an increased floating-point precision where necessary~\cite{colberg2011, ruymgaart2011}. The software 
minimizes disk usage by the \emph{in situ} evaluation of thermodynamic observables and dynamic correlation functions and by 
writing structured, compressed, and portable H5MD output files~\cite{debuyl2014}. Concerning the performance of the package, it 
has been shown to reliably reproduce the slow glassy dynamics of the Kob--Andersen mixture~\cite{colberg2011}, and it was used 
recently to shed new light on the structure of liquid--vapor interfaces~\cite{hoefling2015}.

For the thermalization of the initial state, we have used a Nos\'{e}--Hoover thermostat (NHT) chain ~\cite{frenkel2002, martyna1992} 
with an integration time step of $\delta t=0.002 t_0$. We note that NHT dynamics has recently been demonstrated ~\cite{roy2014} 
to generate critical transport in binary liquid mixtures within the universality class of model~$H'$. We have applied the 
following equilibration procedure: \emph{(i)}~Generate an initial lattice configuration with the desired particle numbers 
$N_\text{A}, N_\text{B}$, and the volume $V$ such that the two species of the particles are randomly assigned and that the 
total momentum is zero. \emph{(ii)}~Melt this lattice at the temperature $2 T$ for $100 t_0$ using the NHT and 
further equilibrate it at~$T$; typical run lengths are $t=10^4 t_0$ for $L=42\sigma$ and $\rho \sigma^3=1.0$. \emph{(iii)}~Determine 
the average internal energy~$U$ at $T$ from the previous NHT run and rescale the particle velocities such that the instantaneous 
total energy matches $U$. The resulting system state is used to compute transport quantities in a production run at fixed total energy, 
employing the velocity Verlet algorithm with an integration time step of $\delta t=0.001 t_0$. For model~II with 
$\varepsilon_{AB}/\varepsilon=1$ we have used $\delta t=0.0005 t_0$. These choices for $\delta t$ result in a relative 
energy drift of less than $2\times10^{-5}$ in $1.5\times10^7$ steps.

\subsection{Data acquisition and statistics}

All results presented in the following correspond to the critical composition $x_c=1/2$ and $T^* \ge T_c^*$. Unless stated 
otherwise, static quantities are all averaged over 20 independent initial configurations and for dynamic quantities this 
number is 30. During the production runs in the NVE ensemble (i.e., $N_A$, $N_B$, $V$, and $E$ constant), data are recorded 
over a time span of $t=15{,}000 \, t_0$. For example, the computing time for a system trajectory of 74,088 particles over 
$1.5\times 10^7$ steps, using a single Tesla K20Xm GPU (NVIDIA Corp.), was 7.1~h at the wall clock, including the evaluation 
of static and dynamic correlations.

\section{Results}
\label{sec:results}

\subsection{Phase diagrams}
\label{subsec:phasedia}

\begin{figure}
\includegraphics[width=0.4\textwidth]{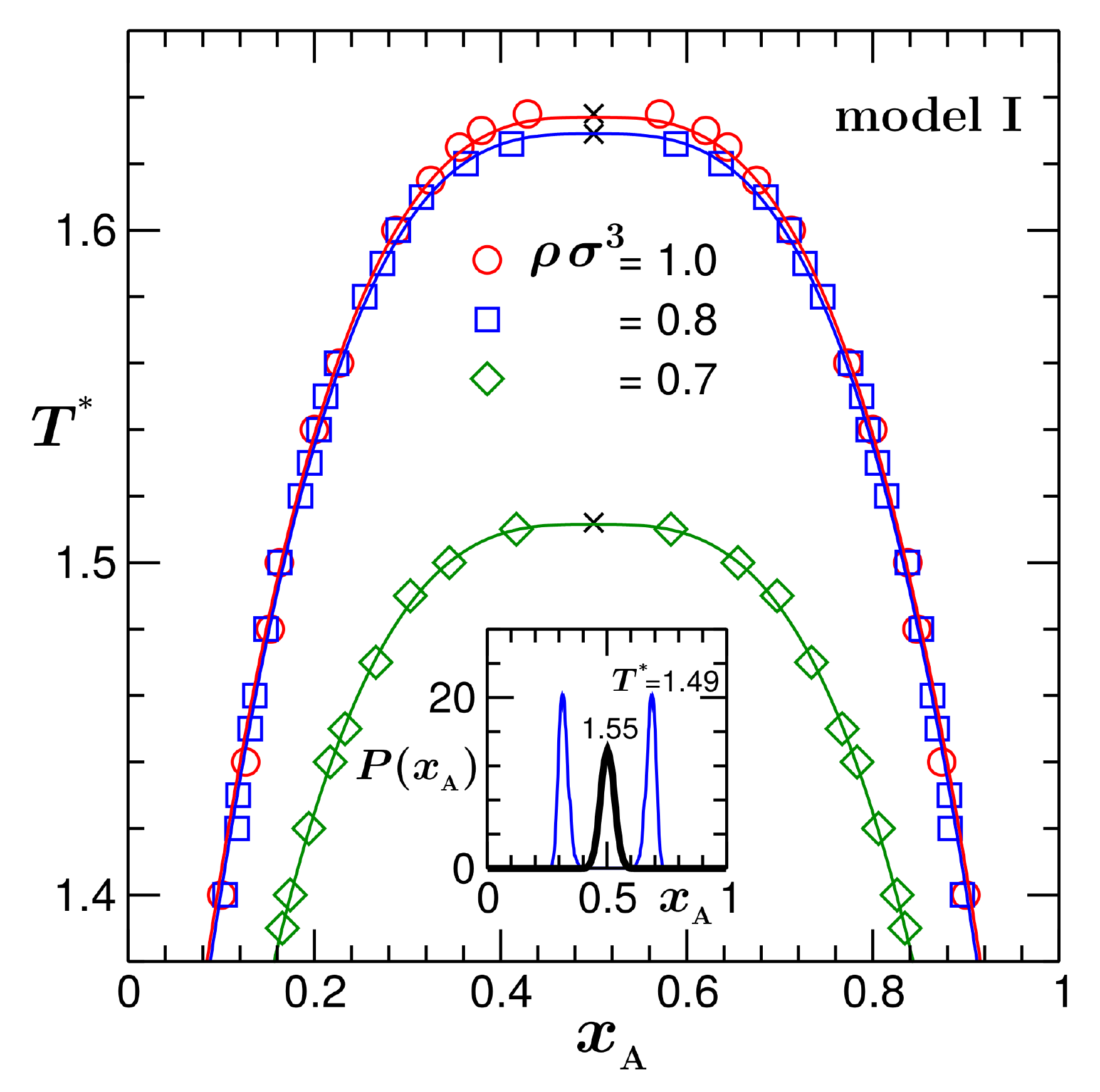}
\caption{Demixing phase diagrams in the $\xA$--$T^*$ plane for binary liquid mixtures within model~I for 3 number densities 
$\rho$ [$\xA=N_A/(N_A+N_B)$ and $T^*=\kB T /\varepsilon$ with $\varepsilon=\varepsilon_{AA}$]. Open symbols indicate 
co-existing equilibrium states obtained from SGMC simulations. Solid lines are fits to these data yielding estimates for 
$T_c(\rho)$ (see the main text), and crosses mark the critical points $(x_c=1/2, T_c)$. The chosen system sizes are 
$L=27\sigma$, $29\sigma$, and $30\sigma$ for $\rho \sigma^3=1.0$, $0.8$, and $0.7$, respectively. In all cases, the 
statistical errors do not exceed the symbol sizes. The inset shows the probability density $P(\xA)$ for model~I at 
$\rho\sigma^3=0.7$, $L=30\sigma$, and two values of $T^*$.}
\label{fig:phasedia}
\end{figure}

From the SGMC simulations we have obtained the demixing phase diagrams for the 5 fluids studied. For each fluid, the probability 
density $P(\xA)$ of the fluctuating concentration $\xA$ of A particles has been determined at various dimensionless 
temperatures $T^*$ above and below the anticipated demixing point; $P(\xA)$ is normalized: $\int_0^1 \! P(\xA) \, \diff\xA=1$. 
In a finite system of linear size~$L$ with periodic boundary conditions along all directions the critical transition 
is shifted and rounded, following the finite-size scaling relation $T_c(L \to \infty) - T_c \sim L^{-1/\nu}$~\cite{fisher1971}. 
In our simulations, we have used large values for $L$ so that the finite-size effects are sufficiently small. The inset of 
\cref{fig:phasedia} shows results for $P(\xA)$ for model~I with $\rho\sigma^3=0.7$ and for two exemplary temperatures: $P(\xA)$ 
shows a single peak above $T_c^L$ and assumes a double-peak structure below $T_c^L$. These peaks of $P(\xA)$ correspond to 
equilibrium states because, up to an $\xA$-independent constant, the free energy is given by 
$-\kB T \log\boldsymbol(P(\xA)\boldsymbol)$. Accordingly, the two peaks of equal height indicate the coexistence of an A- and 
a B-rich phase below $T_c^L$. Due to the symmetry of the binary liquid mixtures, one has $P(\xA)=P(1-\xA)$, which we have imposed 
on the data for~$P(\xA)$. Therefore, the critical composition is $x_{\text{A},c} = x_{\text{B},c} = x_c = 1/2$ which holds 
exactly for all the models considered by us here.

Accordingly, the fluctuating order parameter is given by $\phi = \xA - 1/2$, from which we have calculated the mean order 
parameter $\varphi$ as
\begin{equation}
  \varphi =  \expect{|\phi|} =  \int_0^1 \! \bigl| \xA- 1/2  \bigr| \, P(\xA) \, \diff \xA \,.
\end{equation}
Below $T_c$, where $\varphi > 0$, the binodal is given by the coexisting concentrations $\xA^{(1,2)}(T) = 1/2 \pm \varphi(T)$. 
The results are shown in \cref{fig:phasedia}, and one expects that they follow the asymptotic power law 
\begin{equation}\label{orderparam}
  \varphi(T \nearrow T_c) \simeq \varphi_0  \, |T/T_c - 1|^{~\beta} \,,
\end{equation}

which defines also the amplitude $\varphi_0$. However, deviations are expected to occur for $T$ very close to $T_c$ due to the 
finite-size effects mentioned before~\cite{das2003, roy2011}. For each of the 5 fluids studied, $T_c$ and $\varphi_0$ have been 
estimated via fits of \cref{orderparam} to the data, with $\beta=0.325$ fixed [\cref{staticexponentvalue}].
Ideally, all three parameters $(\varphi_0$, $\beta$, $T_c)$ can be obtained from a single fit procedure as described above. 
However, trying to extract an unknown exponent close to $T_c$ from data for finite-sized systems is a delicate task which usually 
leads to large uncertainties. Already for the extraction of $\varphi_0$ and $T_c$ alone one has to choose the fit range judicially: 
data points very close to $T_c$ suffer from finite-size effects, while the asymptotic law is not expected to hold at temperatures 
far away from $T_c$. Exemplarily for $\rho\sigma^3=0.7$, we have chosen $\xA \in (0.2, 0.4)$. Surprisingly, the power law in 
\cref{orderparam} provides a good description of the binodal even at temperatures well below~$T_c$.

\begin{figure}
\includegraphics[width=0.4\textwidth]{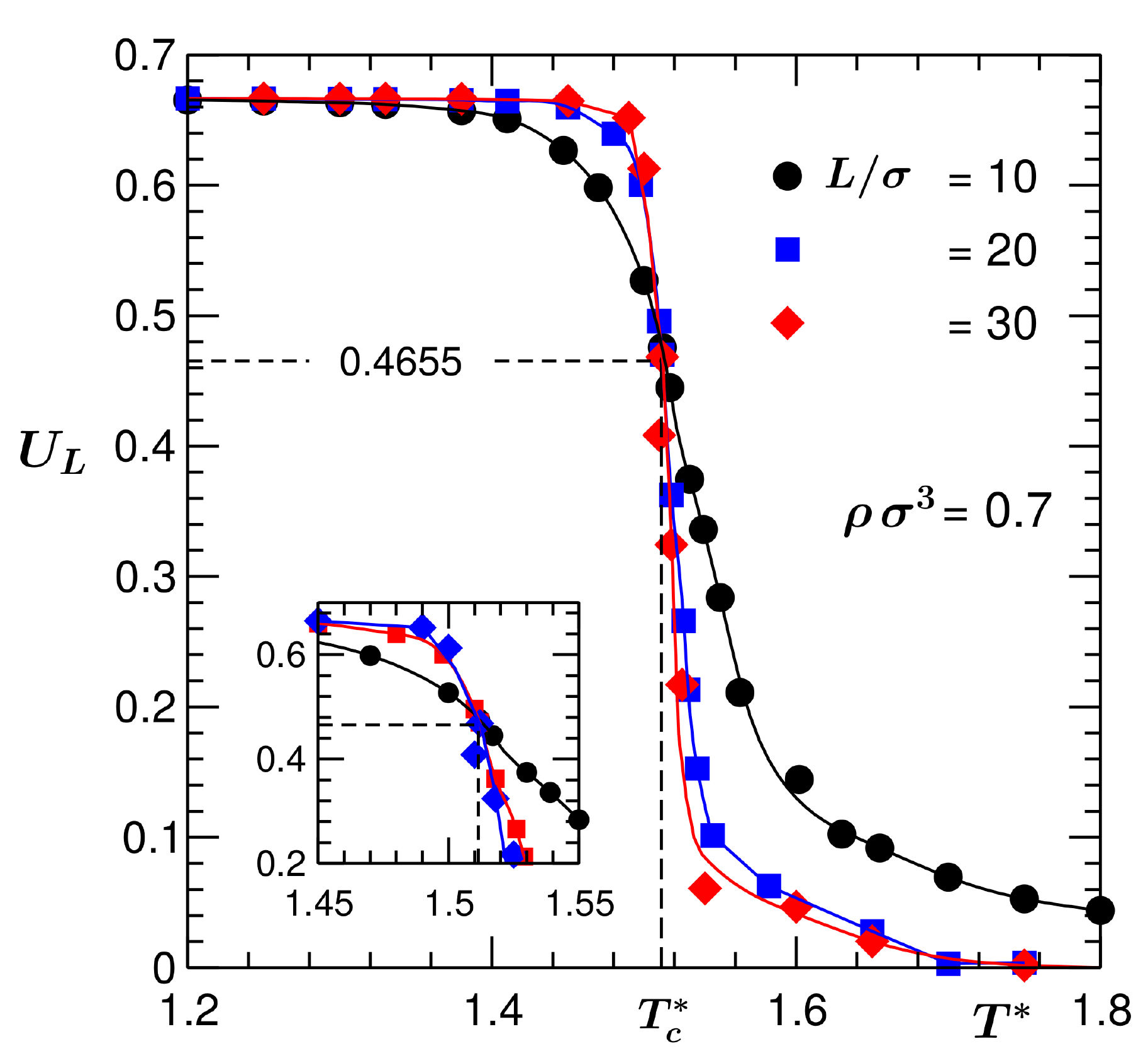}
\caption{Binder cumulant $U_L(T)$ [\cref{bincum1}] for model~I with $\rho\sigma^3=0.7$ and for 3 values of $L$. The solid lines 
are interpolating weighted splines and the dashed lines mark the common intersection point. The inset provides an enlarged view 
of the neighbourhood of the intersection point at $T_c^*$.
}
\label{fig:bincum}
\end{figure}

We have refined the estimates for $T_c$ with Binder's intersection method~\cite{binder1981, binder2010}. It is based on the 
dimensionless cumulant
\begin{equation}\label{bincum1}
  U_L(T)=1-\frac{\expect{ \phi^4} }{3\expect{ \phi^2 }^2} \,,
\end{equation}
which interpolates between the limiting values $U_L(T \to 0) = 2/3$ and $U_L(T \to \infty) = 0$ and, at $T_c$, it attains a 
universal value $U_L(T_c)$ for sufficiently large~$L$. Plotting $U_L(T)$ vs. $T$ for various system sizes~$L$, the set of curves 
exhibits a common point of intersection, from which one infers an accurate estimate for~$T_c$. This is demonstrated in 
\cref{fig:bincum}, showing a family of intersecting curves $U_L(T)$ for model~I with $\rho \sigma^3 = 0.7$ and for three values 
of $L$. The figure corroborates the critical value $U_L(T_c) \approx 0.4655$ \cite{binder1981} for the Ising universality 
class, from which we read off $T_c^*=1.5115\pm 0.0008$.

\begin{table*}
\renewcommand\tabularxcolumn[1]{>{\hfill}p{#1}<{\hfill\hbox{}}}
\heavyrulewidth=.1em
\begin{tabularx}{\textwidth}{c@{\extracolsep{1em}}d{8}d{8}d{8}@{\extracolsep{2em}}d{8}d{8}@{\extracolsep{2em}}d{8}}
\toprule
  & \multicolumn{3}{c}{model I}
  & \multicolumn{2}{c}{model II}
  & \multicolumn{1}{c}{Ref.~\cite{das2006jcp}} \\
\cmidrule{2-4} \cmidrule{5-6} \cmidrule{7-7}
$\rho \sigma^3$ & \multicolumn{1}{X}{0.7} & \multicolumn{1}{X}{0.8} & \multicolumn{1}{X}{1.0} & \multicolumn{1}{X}{0.8} & \multicolumn{1}{X}{0.8} & \multicolumn{1}{X}{1.0} \\
$\varepsilon_{AB} / \varepsilon$ & \multicolumn{1}{c}{1/2} & \multicolumn{1}{c}{1/2} & \multicolumn{1}{c}{1/2} & \multicolumn{1}{c}{1/4} & \multicolumn{1}{c}{1} & \multicolumn{1}{c}{1/2} \\
$r_{c,AB} / \sigma$ & \multicolumn{1}{c}{2.5} & \multicolumn{1}{c}{2.5} & \multicolumn{1}{c}{2.5} &
  \multicolumn{1}{c}{$2^{1/6}$} & \multicolumn{1}{c}{$2^{1/6}$} & \multicolumn{1}{c}{2.5 + force shift} \\
\midrule[\heavyrulewidth]
$T_c^*$ & 1.5115(8) & 1.629(1) & 1.635(3) & 2.608(2) & 4.476(2) & 1.4230(5) \\
$P_c^*$ & 2.58(5) & 4.81(3)& 12.70(4) & 8.29(5) & 16.31(6)& \\
$U_c/\varepsilon$ & -0.617(3)& -0.584(3) & -0.572(4) & 2.643 (4) & 6.557(6) & \\
$\kappa_c^*$ & 0.11(4) & 0.05(3) & 0.02(3) & 0.04(4) & 0.02(4) & \\
\midrule
$\varphi_0$ & 0.76(2) & 0.77(2) & 0.745(16) & 0.77(3) & 0.70(2) & 0.765(25) \\
$\chi_0^*$ & 0.157(9) & 0.112(7) & 0.068(4) & 0.056(4) & 0.06(2) & 0.076(6) \\
$\xi_0/\sigma$ & 0.53(3) & 0.47(2) & 0.42(2) & 0.45(3) & 0.53(2) & 0.395(25) \\
$R_{\xi}^+ \, R_c^{-1/d}$ & 0.71(7) & 0.70(6) & 0.72(6) & 0.72(8) & 0.65(9) & 0.69(8) \\
\midrule
$\eta_0^*$ & & 1.46(10) & 3.63(10) & 1.35(5) & 1.80(7) & 3.87(30) \\
$\mathscr{L}_0^*$ & 0.0143(5) & 0.0082(4) & 0.0024(3) & 0.0049(4) & 0.0032(4) & 0.0028(4) \\
${\mathscr L}_{\text{b},0}^*$ & 0.0082(7) & 0.0040(6) & 0.0028(4) & 0.0025(5) & 0.0013(7) & 0.0033(8) \\
$D_{m,0}^*$ & 0.091(8) & 0.073(6) & 0.035(5) & 0.088(7) & 0.053(6) & 0.037(8) \\
$R_D$ & & 0.95(17) & 0.993(21) & 1.00(21) & 0.96(25) & 1.06(40) \\
\bottomrule
\end{tabularx}
\caption{Simulation results for the five binary fluids investigated here along with results from Ref.~\cite{das2006jcp}: values 
of the critical temperature~$T_c$ and of the pressure~$P_c$, the internal energy~$U_c$ [\cref{energyeq}], and the isothermal 
compressibility~$\kappa_c$ at the critical point; further, the critical amplitudes of the correlation length~$\xi_0$ [\cref{OZ}], 
of the order parameter~$\varphi_0$, and of the static susceptibility~$\chi_0$ [\cref{OZ}], as well as of the shear viscosity~$\eta_0$ 
[\cref{shearGK,shearGK2}] and the Onsager coefficient $\mathscr{L}_0$ [\cref{onsdef,ons3,ons4}]. The amplitudes $D_{m,0}$ 
of the interdiffusion constant have been computed from \cref{eq:DAB_amplitude}. Finally, the values of two universal ratios 
of static and dynamic amplitudes, $R_\xi^+ R_c^{-1/d}$ and $R_D$ [see \cref{twoscaleuniv,dynunivratio}, respectively], are reported. 
Numbers in parentheses indicate the uncertainty in the last digit(s).
}
\label{tab:results}
\end{table*}

The main results for all 5 fluids studied here have been compiled in \cref{tab:results}. For model~II, we have found much 
higher values for $T_c$ than for model~I. This can be understood as follows. The AA and BB interaction potentials are 
identical both within and between the two models, which differ only with respect to the AB interaction potential. Upon 
construction [see \cref{LJ2}], the AB interaction is more repulsive in model~II than in model~I ($u_\text{AB}^\text{II}$ is 
purely repulsive whereas $u_\text{AB}^\text{I}$ exhibits also an attractive part). A repulsive AB interaction favors the 
formation of domains rich in A and of domains rich in B and thus promotes demixing, which leads to a higher value of $T_c$. 
Within model~II, increasing $\varepsilon_\text{AB}$ makes $u_\text{AB}^\text{II}$ more repulsive and thus renders the same 
trend. In particular, increasing within model~II the attraction strength $\varepsilon_\text{AB}$ by a factor of~4 yields 
an $1.7$-fold increase of~$T_c$. The binodal for model~II with $\varepsilon_\text{AB}/\varepsilon=1$ (but $u_{\alpha\alpha}$ 
truncated at $r_c/\sigma =4.2$) was determined in Ref.~\cite{toxvaerd1995}, and the rough estimate of $T_c$ there agrees 
with our result. Model~I was studied by \textcite{das2003} for $\rho \sigma^3 =1.0$, but using sharply truncated interaction 
potentials [$f(x)\equiv \theta(-x)$ in \cref{LJ2}]. They found $T_c^* = 1.638 \pm 0.005$, which is very close to our result 
$T_c^* =1.635 \pm 0.003$, however different from $T_c^*\approx 1.423$ as obtained for the force-shifted potentials used in 
Ref.~\cite{das2006jcp}. We propose that this difference appears because the smoothing function $f(x)$ alters $u_{\alpha\beta}(r)$ 
only locally near $r_c$, unlike the force shift which amounts to modify the interaction potentials globally. Within model~I, 
the amplitude $\varphi_0$ of the order parameter is, within the accuracy of our data, insensitive to the density $\rho$ (see 
\cref{tab:results}). Within model~II, $\varphi_0$ changes only slightly upon increasing the strength~$\varepsilon_\text{AB}$ of the 
repulsion.

\begin{figure}
\centering
\includegraphics*[width=0.4\textwidth]{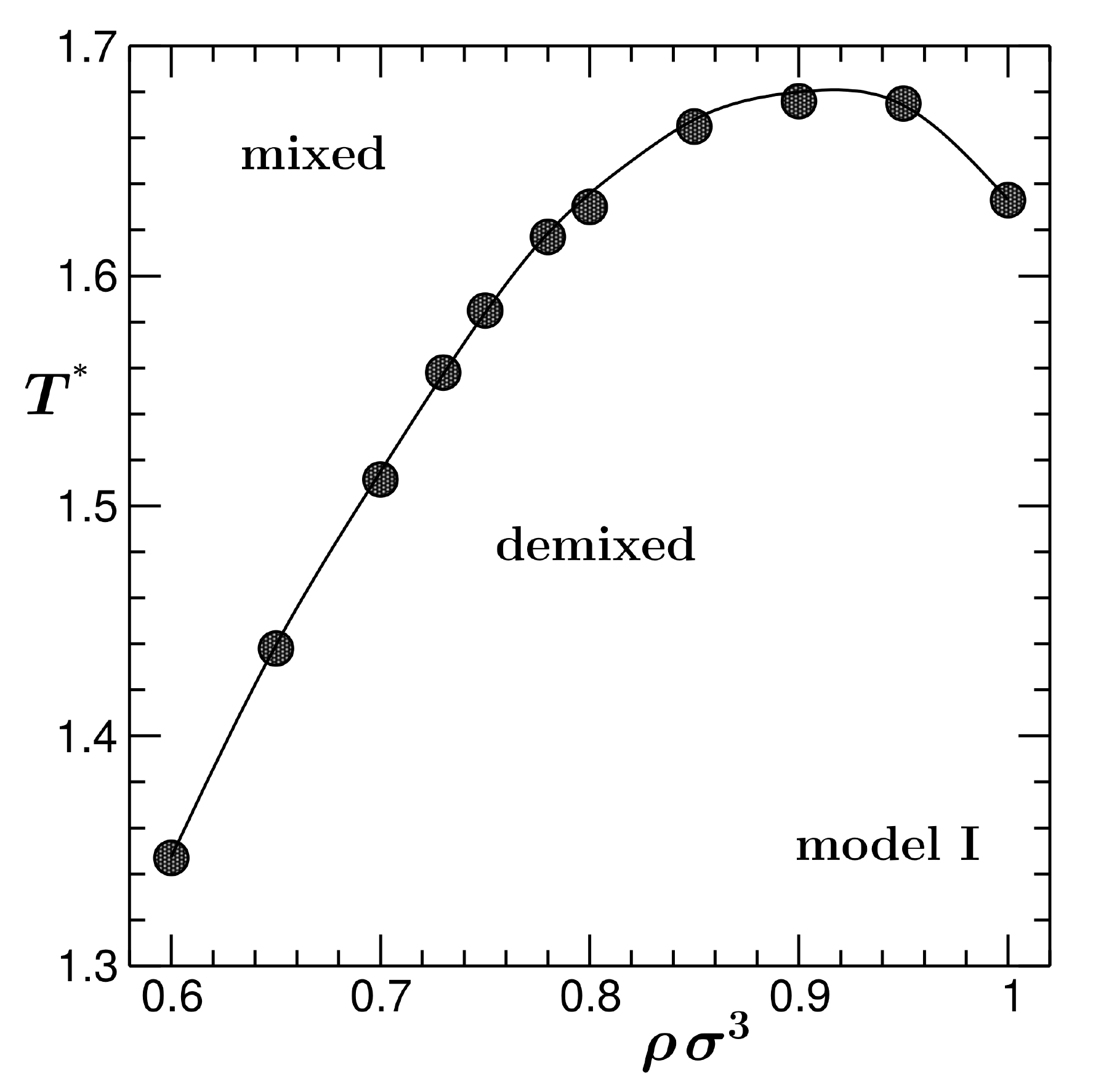}
\caption{Loci of the critical points $T_c(\rho)$ for the demixing transition in the $\rho$--$T^{*}$ plane with $T^*=\kB T /\varepsilon$, 
obtained from SGMC simulations for model~I. The solid line is an interpolating weighted spline. For details of the interaction potentials, 
see the main text and \cref{tab:results}. Error bars are smaller than the symbol sizes.}
\label{fig:lambdaline}
\end{figure}

The dependence of the demixing transition on the total number density $\rho$ gives rise to a line $T_c(\rho)$ of critical 
points, known as the \emph{$\lambda$-line} (\cref{fig:lambdaline}). For model~I, we have found that $T_c(\rho)$ is an increasing 
function for $\rho\sigma^3 \lesssim 0.9$; for higher densities, it decreases. Such a non-monotonic dependence implies a 
re-entrance phenomenon: increasing the density isothermally, the binary liquid mixture undergoes a phase transition from 
a mixed state at low density to a phase-separated one and mixes again at high densities.

The initial increase of $T_c(\rho)$ is in qualitative agreement with a previous grand-canonical MC study~\cite{wilding1997} 
using a variant of our model~I ($\varepsilon_\text{AB}=0.7 \varepsilon_\text{AA}$). For this choice of the interaction 
potentials, it was found that the $\lambda$-line ends at a critical end point near $\rho \sigma^3 \approx 0.59$, where the 
$\lambda$-line hits the first-order liquid--vapor transition of the fluid. It was suggested \cite{wilding1997} that upon 
decreasing $\varepsilon_\text{AB}$ further the critical end point moves towards the line of liquid--vapor critical points 
until both lines of critical points meet for $\varepsilon_\text{AB} \approx 0.6 \varepsilon_\text{AA}$ and form a tri-critical 
point, as observed in a two-dimensional spin model~\cite{wilding1996}. The determination of the full phase diagrams of binary 
liquid mixtures, encompassing the complete $\lambda$-line, the line of liquid--vapor critical points, and the solid phases, 
is a non-trivial and computationally demanding task (for density functional approaches in this direction see Refs. 
\cite{dietrich1989, getta1993, dietrich1997}). It remains as an open question whether there is a tri-critical point in model~I 
($\varepsilon_\text{AB}=0.5 \varepsilon_\text{AA}$) or not (see below for further discussions).

For completeness, \cref{tab:results} also lists the critical values for the dimensionless pressure $P_c^*=P_c \sigma^3/\varepsilon$ 
at the respective demixing points, which have been obtained, following the standard procedures for a homogeneous and isotropic 
fluid, from the trace of the time-averaged stress tensor, $P = \tr \expect{ \Pi(t) } /3 V$, with the instantaneous stress 
tensor given by~\cite{hansen2008}
\begin{align}\label{stresstensor}
\Pi(t)&=\sum_{i=1}^N \Biggl\{
  m \vec v_i(t) \otimes \vec v_i(t) +
  \sum_{j > i}^N \vec r_{ij}(t) \otimes \vec F_{ij}(t)
\Biggr\} \,,
\end{align}
where $\vec v_i(t)$ is the velocity of particle $i$, $\vec r_{ij}(t) = \vec r_i(t) - \vec r_j(t)$, $\vec F_{ij}(t)$ is the force 
acting on particle~$j$ due to particle~$i$, and $\otimes$ denotes a tensor product. We also specify the internal energy per particle,
\begin{equation}\label{energyeq}
  U = \frac{1}{N} \sum_{i=1}^{N} \biggl\langle\frac{m}{2} \vec v_i(t)^2
  + \sum_{j>i}^N u_{\alpha_i\alpha_j}\bigl(|\vec r_{ij}(t)|\bigr) \biggr\rangle \,,
\end{equation}
at the critical points, which characterizes the microcanonical ensemble probed by MD simulations; $\alpha_i \in \{A,B\}$ 
denotes the species of particle~$i$.

\subsection{Spatial correlations}

\subsubsection{Static structure factors}

The structural properties of binary liquid mixtures arise from the two fluctuating fields, given by the microscopic partial number 
densities $\rho_\alpha(\vec r)$ of each species~$\alpha$. Their fluctuating parts are~\cite{hansen2008}
\begin{equation}
  \delta \rho_\alpha(\vec r) =  - \frac{N_\alpha}{V} +\sum_{j =1}^{N_\alpha} \delta \mleft(\vec r -\vec r^{(\alpha)}_j \mright) \,,
\end{equation}
where the set $\bigl\{ \vec r^{(\alpha)}_j \bigr\}$ denotes the positions of the $N_\alpha$ particles of species~$\alpha$. It is 
favorable to (approximately~\cite{das2003}) decouple the spatial fluctuations into the overall density contribution 
$\delta \rho(\vec {r})$ and into the composition contribution $\delta c(\vec r)$ and, accordingly, to consider the linear 
combinations~\cite{bhatia1970}
\begin{subequations}\label{fluctuating_fields}
\begin{equation}
  \delta \rho(\vec r) = \delta \rho_\text{A}(\vec r) +  \delta \rho_\text{B}(\vec r) \, \\
  \end{equation}
and
\begin{equation}\label{delta_c}
  \delta c(\vec r) = \xB \delta \rho_\text{A}(\vec r)-\xA \delta \rho_\text{B}(\vec r) \,;
\end{equation}
\end{subequations}
$\delta \rho(\vec r)$ fluctuates around the total density $\rho = (N_\text{A} + N_\text{B}) / V$. In Fourier space, the 
corresponding spatial correlation functions are defined as
\begin{subequations}\label{structurefactors}
\begin{align}
  S_{\rho\rho}(|\vec k|) & = \frac{1}{N} \expect{ \delta \rho_{\vec k}^* \, \delta \rho_{\vec k}}, \\
  S_{cc}(|\vec k|) & = N \expect{ \delta c_{\vec k}^* \, \delta c_{\vec k} },
\intertext{and}
  S_{\rho c}(|\vec k|) & = \Real \expect{ \delta \rho_{\vec k}^* \, \delta c_{\vec k} } ,
\end{align}
\end{subequations}
where, e.g., $\delta \rho_{\vec k} = \int_V \! \e^{\i \vec k \cdot \vec r} \delta \rho(\vec r) \,\diff^3 r$ and, see below, 
$\delta \rho_{\vec k}^{(\alpha)} = \int_V \! \e^{\i \vec k \cdot \vec r} \delta \rho_\alpha(\vec r) \,\diff^3 r$.

On the fly of the MD simulations, we have determined the partial structure factors
$
  S_{\alpha\beta}(|\vec k|)=(f_{\alpha\beta} / N)
  \expect{\delta \rho^{(\alpha)}_{-\vec k} \, \delta \rho^{(\beta)}_{\vec k} },
$
where $f_{\alpha\beta} = 1$ for $\alpha=\beta$ and $f_{\alpha\beta} = 1/2$ for $\alpha \ne \beta$~\cite{hansen2008}, which 
allow one to determine
\begin{subequations}
\begin{align}
S_{\rho\rho}(k) &= S_\text{AA}(k)+S_\text{BB}(k)+2 S_\text{AB}(k) \,, \\
S_{cc}(k) &= \xB^2 S_\text{AA}(k)+\xA^2 S_\text{BB}(k) - 2\xA\xB S_\text{AB}(k) \,,
\intertext{and}
S_{\rho c}(k) &= \xB S_\text{AA}(k)-\xA S_\text{BB}(k) +(\xB-\xA) \, S_\text{AB}(k) \, .
\end{align}
\end{subequations}

\begin{figure}
\centering
\includegraphics*[width=0.4\textwidth]{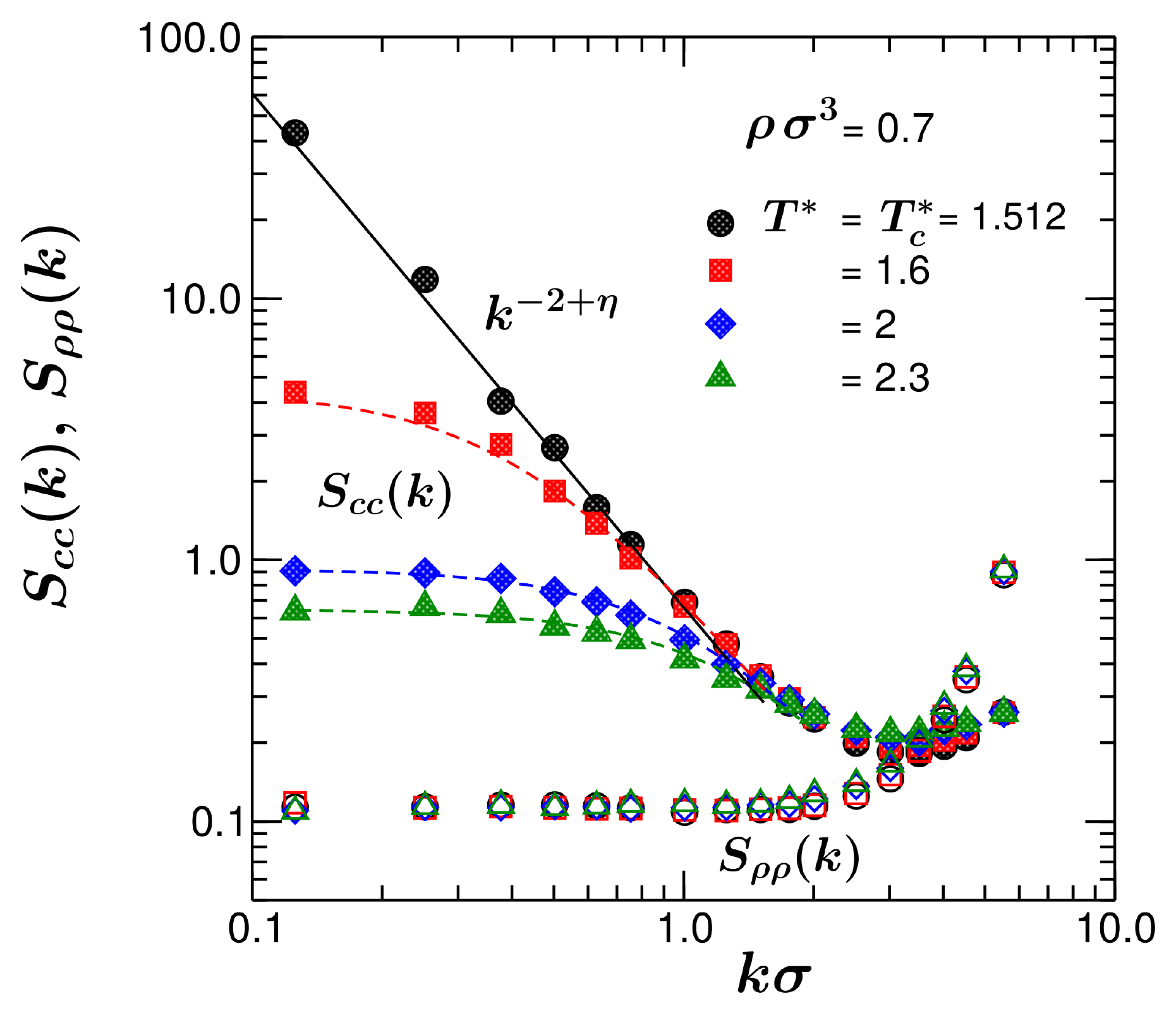}
\caption{Static structure factors $S_{cc}(k)$ (filled upper symbols) and $S_{\rho\rho}(k)$ (open lower symbols) for model~I 
($\rho\sigma^3=0.7$, $L=50\sigma$) at four dimensionless temperatures $T^* \geq T_c^*=1.512$ [see \cref{structurefactors}]. The dashed 
lines show fits to the extended Ornstein--Zernike form [\cref{OZ}] for $T > T_c$. The solid line indicates the critical law 
$S_{cc}(k,T=T_c) \sim k^{-2+\eta}, \: \eta=0.036$, which appears as a straight line on the double-logarithmic scales. 
Error bars are of the size of the symbols.
}
\label{fig:scc}
\end{figure}

Here, we are primarily interested in the critical fluctuations of the composition, which are borne out by $S_{cc}(k)$ for small $k$. 
The latter is of the (extended) Ornstein--Zernike form
~\cite{stanley1971, hansen2008}
\begin{equation}\label{OZ}
  S_{cc}(k) \simeq \frac{\rho \kB T \chi}{[1+k^2\xi^2]^{1-\eta/2}} \,, \quad k\sigma \ll 1 \,,
\end{equation}
which defines both the static order parameter susceptibility $\chi \sim \tau ^{-\gamma}$ and the correlation length 
$\xi \sim \tau^{-\nu}$ which in real space governs the exponential decay of the correlation functions. The anomalous dimension 
$\eta \approx 0.036$ follows from the exponent relation $\gamma=\nu(2-\eta)$~\cite{pelissetto2002} and the values in 
\cref{staticexponentvalue}. Exemplary results for $S_{cc}(k)$ and $S_{\rho\rho}(k)$ are shown in \cref{fig:scc}, for model~I 
with $\rho\sigma^3=0.7$, on double-logarithmic scales. For the studied range of temperatures, $T_c^* \leq T^* \leq 2.3$, all curves 
for $S_{cc}(k)$  display a minimum near $k\sigma \approx 3$ and are not sensitive to temperature for $k$ larger than this. 
$S_{cc}(k)$ increases as $k \to 0$, which, due to the divergence of $\chi \sim \tau^{-\gamma}$ at $T_c$, becomes stronger as 
$T\to T_c$. This reflects the enhancement of the critical composition fluctuations with a concomitant increase of the 
correlation length~$\xi$. The data for $S_{cc}(k)$ for $T>T_c$ exhibit a nice consistency with the theoretical extended 
Ornstein--Zernike form, depicted by the dashed lines in \cref{fig:scc}, over approximately one decade in wavenumber~$k$. 
Right at $T_c$, the data for $S_{cc}(k)$ follow for one decade in $k$ the expected critical power law~\cite{stanley1971}
\begin{equation}\label{OZ2}
    S_{cc}(k \to 0, T=T_c) \sim k^{-2+\eta} \,,
\end{equation}
emerging from \cref{OZ} for $k\xi \gg 1$.

To the contrary, the density fluctuations, described by $S_{\rho\rho}(k)$, do not change appreciably within the temperature 
range considered (\cref{fig:scc}). There is no critical enhancement for small wave numbers and the spatial range of the 
density--density correlations is short near the demixing transition with $\rho\sigma^3=0.7$. The value of 
$S_{\rho\rho}(k\to 0)=\rho \kB T \kappa_T$ yields the isothermal compressibility $\kappa_T=\kappa_T^*\sigma^3/\varepsilon$ 
of the fluid, which diverges at a liquid--vapor critical point. Along the $\lambda$-line $T_c(\rho)$, it increases from 
$\kappa_c^* := \kappa_T^*(T=T_c) \approx 0.02$ for the highly incompressible fluid at $\rho\sigma^3 = 1$ to 
$\kappa_c^* \approx 0.11$ at $\rho\sigma^3 = 0.7$ (\cref{tab:results}). In all cases one has 
$S_{\rho\rho}(k\to 0) \ll S_{cc}(k\to 0)$. From this we conclude that, for the densities considered here, the liquid--vapor and the 
demixing critical points are sufficiently well separated.

\begin{figure}
\centering
\includegraphics*[width=0.4\textwidth]{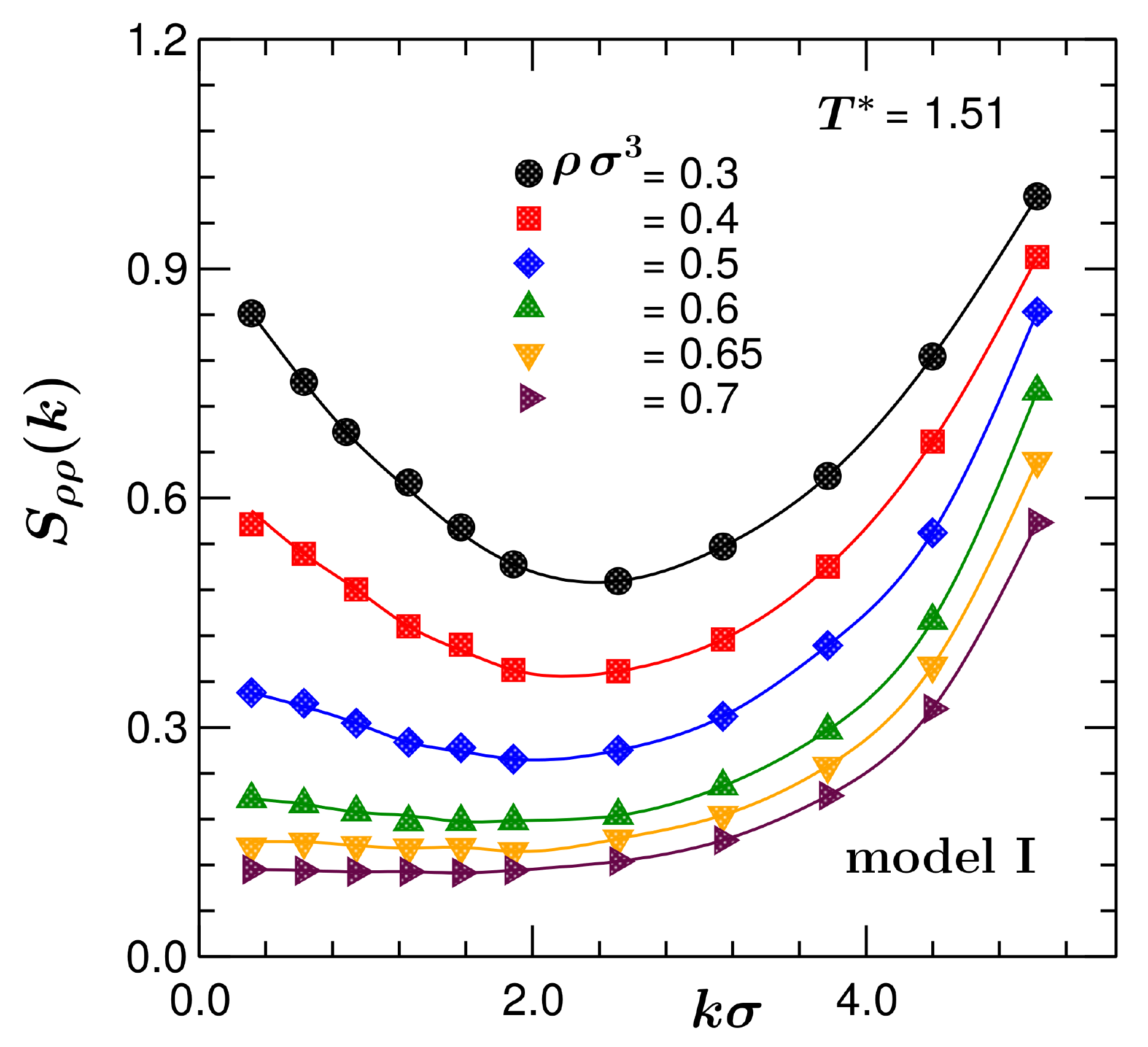}
\caption{Static structure factor $S_{\rho\rho}(k)$ for model~I ($L=20\sigma$) for six number densities~$\rho$ and fixed 
dimensionless temperature $T^*=1.51$. Lines are drawn by using suitable splines. }
\label{fig:snn_allrho}
\end{figure}

In order to probe the location of the line of the liquid--vapor critical points, we have lowered the density along the isotherm 
$T^*=1.51 \approx T_c^*(\rho\sigma^3=0.7)$. Indeed, for $\rho\sigma^3 \lesssim 0.6$ the corresponding structure factors 
$S_{\rho\rho}(k)$, shown in \cref{fig:snn_allrho}, display the emergence of critical density fluctuations via a monotonic 
increase of the compressibility by a factor of~19. Further, the value of $\kappa_T^* (\rho\sigma^3=0.3,T^*=1.51)$ is ca. 7.4 times 
larger than $\kappa_c^*\simeq 0.25$ at $T_c(\rho\sigma^3=0.6)$, following the $\lambda$-line (\cref{fig:lambdaline}). This suggests 
that $\kappa_T^*$ does not diverge along the $\lambda$-line, which implies that the $\lambda$- line and the line of 
liquid--vapor critical points do not meet, thus rendering the occurrence of a tri-critical point in model~I as to be unlikely.

\subsubsection{Correlation length and static order parameter susceptibility}

\begin{figure}
\centering
\includegraphics*[width=0.4\textwidth]{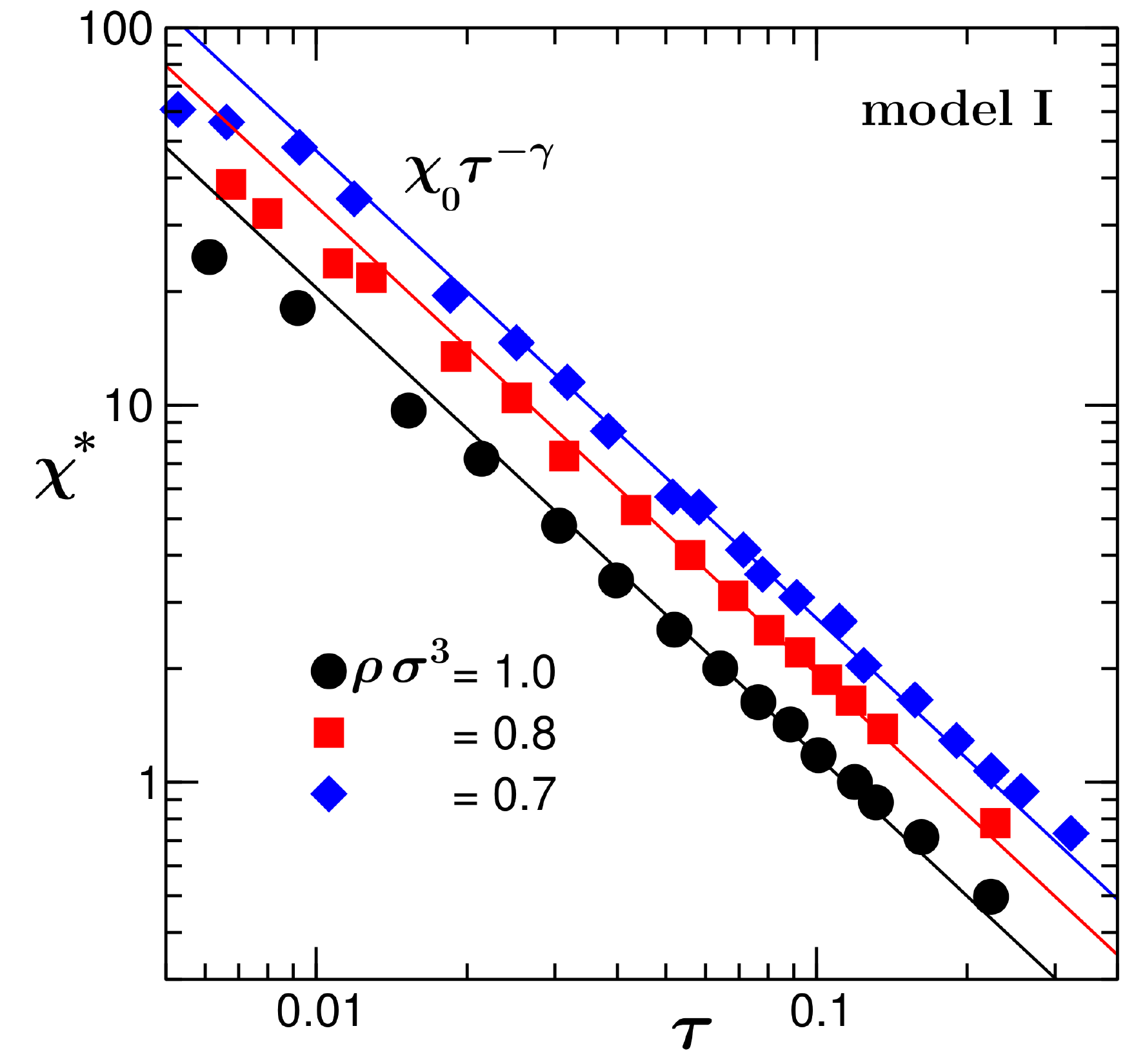}
\caption{Reduced static order parameter susceptibility $\chi^*=\chi \varepsilon \sigma^{-3}$ [\cref{OZ,criticallawsus}] as 
function of $\tau=(T-T_c)/T_c$ within model~I for three number densities. The system sizes are $L/\sigma=42, 47,$ and $50$ for 
$\rho \sigma^3=1.0, 0.8,$ and $0.7$, respectively. Straight lines indicate the asymptotic power law 
$\chi(T \to T_c) \simeq \chi_0 \tau^{-\gamma}$ with $\gamma=1.239$. Error bars are smaller than the symbol sizes. }

\label{fig:sus}
\end{figure}

\begin{figure}
\centering
\includegraphics*[width=0.4\textwidth]{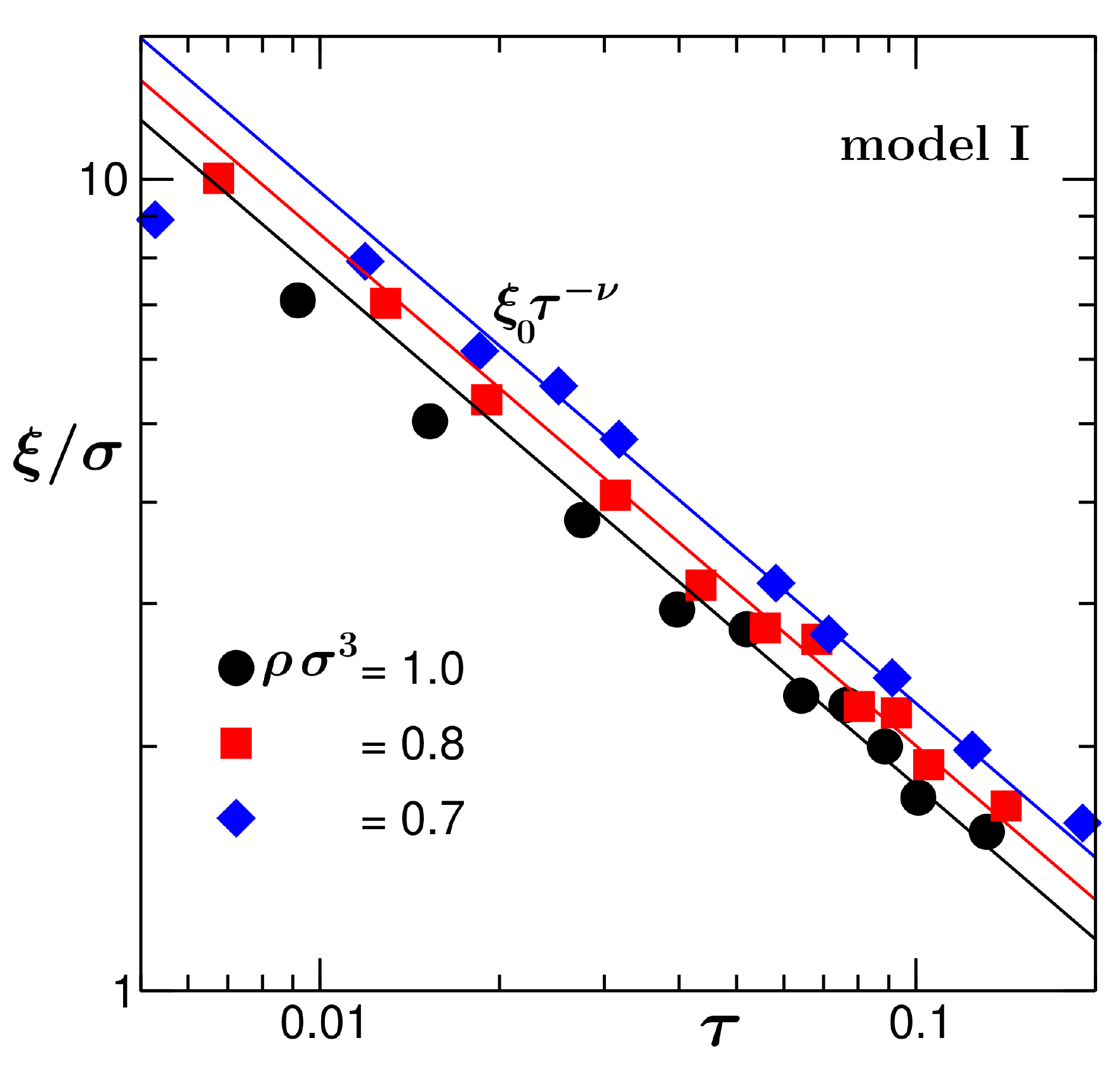}
\caption{Correlation length $\xi$ of the concentration fluctuations [\cref{OZ,criticallawsus}] as function of $\tau$ on 
double-logarithmic scales. Straight lines refer to the asymptotic power law $\xi(T\to T_c) \simeq \xi_0 \tau^{-\nu}$ with 
$\nu=0.630$. All simulation parameters are the same as in \cref{fig:sus}. Relative error bars are within $1-9\%$.}
\label{fig:cor}
\end{figure}

For a broad range of temperatures $T_c \leq T \lesssim 2 T_c$, we have run extensive MD simulations for three binary 
liquid mixtures in model~I and two binary liquid mixtures in model~II. The use of large system sizes has enabled us to 
reach the critical point as close as $\tau = (T-T_c) / T_c \simeq 0.01$. Fitting \cref{OZ} to the data for $S_{cc}(k)$ we have 
obtained the static order parameter susceptibilities $\chi(T)$ and the correlation lengths $\xi(T)$. The data nicely follow 
the asymptotic power laws
\begin{equation}\label{criticallawsus}
  \chi(T \to T_c) \simeq \chi_0^{^{~}} \tau^{-\gamma}\, , \qquad  
  \xi(T\to T_c) \simeq \xi_0^{^{~}} \tau^{-\nu}
\end{equation}

near the respective critical temperatures $T_c$ with the Ising critical exponents $\gamma$ and $\nu$ in $d=3$ (see 
\cref{fig:sus,fig:cor} for the binary liquid mixtures in model~I). Finite-size effects become apparent for $\tau \lesssim 0.02$, 
for which the data for both quantities fall short of the asymptotic law. Interestingly, this occurs already for correlation 
lengths $\xi \approx 4\sigma \approx L/10$. The fit has also identified a temperature range, where corrections to the asymptotic 
power laws are not yet important. In \cref{fig:sus}, the upward trend in $\chi(\tau)$ for $\rho \sigma^3=0.7$ and $\tau>0.2$ 
reflects the necessity for such corrections. The amplitudes $\chi_0$ and $\xi_0$ are non-universal quantities and are listed 
in \cref{tab:results} for each fluid. The trend of a decreasing $\chi_0$ upon increasing $\rho$ (model~I) may be explained by 
the fact that the re-arrangement of particle positions becomes more costly (in terms of potential energy at denser packing, 
reducing the response of the system. Even smaller values of $\chi_0$ have been found for model~II, with no pronounced dependence 
on the strength~$\varepsilon_\text{AB}$ of the repulsion. Across all five binary mixtures the amplitude $\xi_0$ of the correlation 
length varies only mildly between $0.42\sigma$ and $0.53\sigma$.

For comparison, we have also determined the order parameter susceptibility $\chi$ via the SGMC simulations above $T_c$ from the 
variance of the fluctuating composition~\cite{stanley1971, hansen2008}. For the symmetric binary liquid mixtures as considered 
here, one has
$
  \rho \kB T \chi = N \bigl(\expect{ \xA^2 } - \expect{ \xA }^2 \bigr)
$
with $\expect{ \xA } = 1/2$ at $T = T_c$ due to the model symmetry and above $T_c$ for the mixed phase. We have found that the 
results obtained from these two approaches agree.

\subsection{Transport coefficients}

\subsubsection{Interdiffusion constant}

A critical point leaves its marks both in space and time: upon approaching criticality the correlation length diverges and the 
relaxation of a fluctuation or of a perturbation slows down. The latter manifests itself in terms of universal power-law behaviors 
of transport coefficients upon approaching $T_c$. For example, a gradient in the composition field $\delta c(\vec r, t)$
[\cref{delta_c}] generates a collective current~\cite{zhou1996, das2006jcp, horbach2007}
\begin{equation}\label{ons2}
  {\vec J}_\text{AB}(t)=\xB \sum_{i=1}^{N_A} \vec v^{(A)}_{i}(t)
  -\xA \sum_{i=1}^{N_B} \vec v^{(B)}_{i}(t) \,,
\end{equation}
the magnitude of which is captured by the interdiffusion constant~$\Dm$. This coefficient controls the collective diffusion of 
the composition field and obeys a Green--Kubo relation
~\cite{zhou1996, das2006jcp, horbach2007}:
\begin{equation}\label{ons1}
  \Dm = \frac {1}{d N S_{cc}(k=0)}
  \int_0^\infty \! \expect{ \vec J_\text{AB}(t) \cdot \vec J_\text{AB}(0) } \diff t \,,
\end{equation}
where $d$ is the spatial dimension and $S_{cc}(k=0)=\rho \kB T \chi$. The interdiffusion constant is a combination of a static 
property, i.e., the concentration susceptibility $\chi$, and a pure dynamic quantity, the concentration conductivity or Onsager 
coefficient~\cite{zhou1996, das2006jcp, horbach2007}
\begin{equation}\label{onsdef}
  \mathscr{L} = \chi \Dm\,.
\end{equation}
$\mathscr L$ connects gradients in the chemical potentials with the current $\vec J_\text{AB}$; as its dimensionless form we use 
$\mathscr{L}^* := \mathscr{L} \varepsilon t_0 \sigma^{-5}$.

The numerical evaluation of the time integral in \cref{ons1} is challenged by statistical noise and by hydrodynamic long-time tails 
of the current correlators~\cite{hansen2008}. An alternative route to compute $\mathscr L$ is based on the generalized Einstein 
relation~\cite{horbach2007, roy2013, hoeft2012}
\begin{equation}\label{ons3}
  {\mathscr L} = \lim_{t \to \infty} \frac{ N_\alpha^2}{2 \rho N \kB T}
  \frac{\diff}{ \diff t} \,  \delta\mathscr{R}^2_{\alpha}(t)
\end{equation}
for $\alpha\in \{\text{A}, \text{B}\}$ with the collective mean-square displacement
\begin{equation}\label{ons4}
  \delta\mathscr{R}^2_{\alpha}(t) = \expect{ \mleft|\int_0^t \! \vec V_\alpha(t) \, \diff t \, \mright|^2 } \,,
  \quad \vec V_{\alpha}(t) = \frac{1}{N_\alpha} \sum_{i=1}^{N_\alpha} \vec v_{i}^{(\alpha)} (t) \,,
\end{equation}
defined in terms of the centre-of-mass velocity $\vec V_{\alpha}(t)$ by considering particles of species~$\alpha$ only. Note that 
${\vec V}_\text{A}(t) = -{\vec V}_\text{B}(t)$ for a symmetric mixture ($m_A=m_B$, $N_\text{A} = N_\text{B}$) due to conservation 
of the total momentum, $\sum_\alpha m_\alpha N_\alpha \vec V_\alpha = 0$. Our simulation data tell that the results for $\mathscr{L}$ 
as obtained from both methods [\cref{ons1,ons3}] coincide within the error bars. The latter route, however, exhibits superior 
averaging properties, in line with previous findings for a different system concerning the motion of a tagged particle~\cite{hoefling2007}. 
The success of the method hinges on evaluating $\delta \mathscr{R}_\alpha^2(t)$ by using a certain ``blocking scheme''
~\cite{frenkel2002, colberg2011}, which resembles a non-averaging multiple-$\tau$ correlator and naturally generates a 
semi-logarithmic time grid, particularly suitable for the description of slow processes. With this, the time derivative 
in \cref{ons3} can simply be computed from central difference quotients. The results for $\mathscr L$ presented here have been 
obtained by applying this method.

\begin{figure}
\centering
\includegraphics*[width=0.4\textwidth]{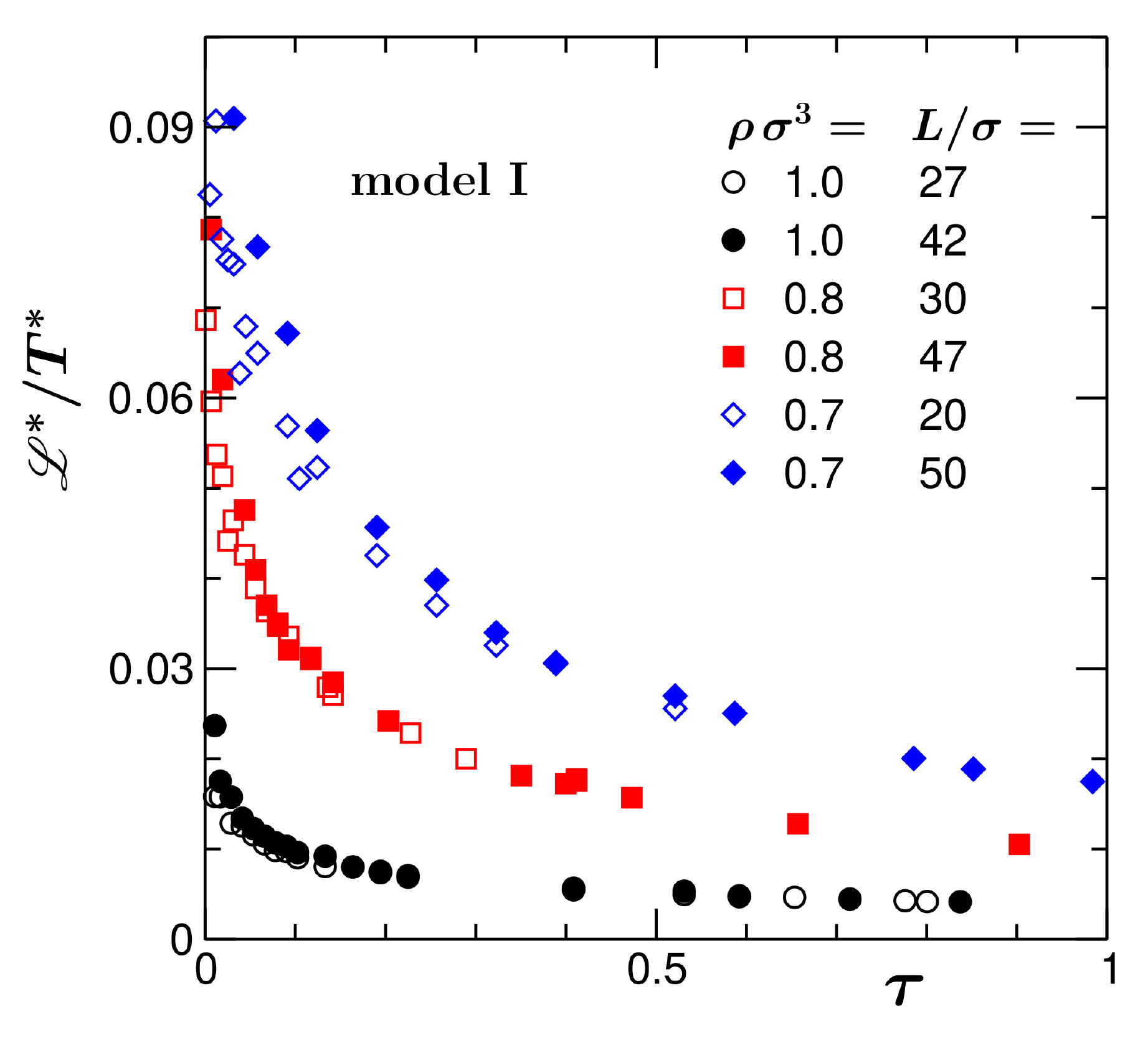}
\caption{Plot of $\mathscr L^*/T^*$ with $\mathscr L^* = \mathscr{L} \varepsilon t_0 \sigma^{-5}$ [\cref{ons1,onsdef,ons3,ons4}]
for model~I as function of the reduced temperature $\tau$, for three number densities~$\rho$. For each~$\rho$, data for two 
different system sizes $L$ are presented. Relative error bars are within 3--9\%.
}
\label{fig:ons_different_density}
\end{figure}

\begin{figure}
\centering
\includegraphics*[width=0.4\textwidth]{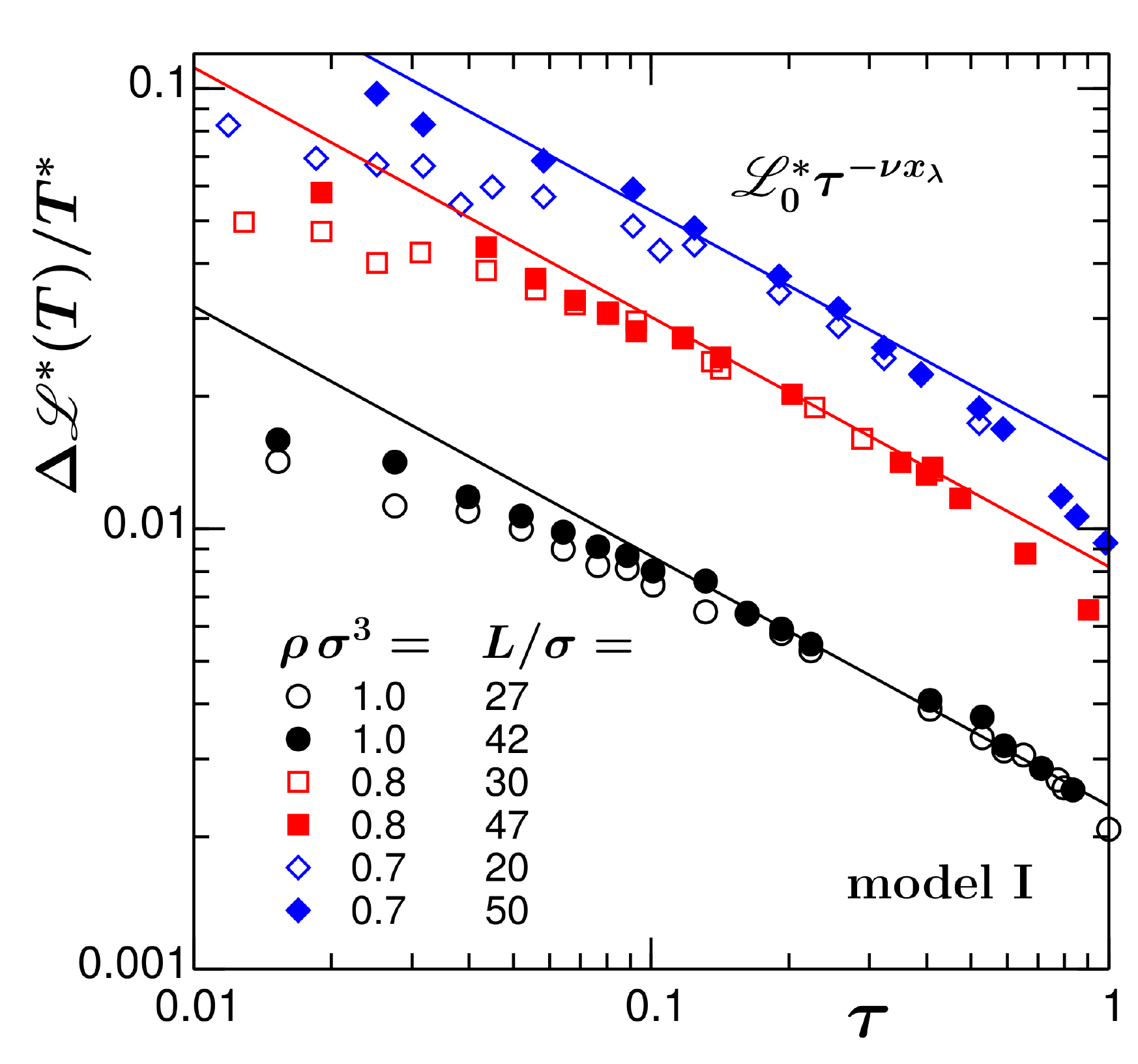}
\caption{The same data as in \cref{fig:ons_different_density} in terms of
$\Delta \mathscr{L}^*/T^* = \mathscr{L}^*/T^* - \mathscr{L}_{\text{b},0}^*$ after adjusting the background contribution 
$\mathscr{L}_{\text{b},0}^*$. Solid lines refer to the power law 
$\Delta {\mathscr L}/(\kB T) \simeq \mathscr{L}_0 \tau^{-\nu x_\lambda}$ with $\nu x_\lambda =0.567$ [\cref{ons5}].
}
\label{fig:ons_quantification}
\end{figure}

The interdiffusion constant [\cref{ons1}] can be decomposed as $\Dm=\Delta\Dm + D_b$ into a singular contribution $\Delta \Dm$ 
stemming from critical fluctuations in the fluid at large length scales and an omnipresent analytic background term $D_b$ 
arising due to short-length-scale fluctuations \cite{sengers1985}. As predicted by MCT and dynamic RGT, asymptotically 
close to the critical temperature $\Delta \Dm$ follows the Einstein--Kawasaki relation \cite{kawasaki1972,onuki1997}:
\begin{equation}\label{DeltaDAB}
\Delta \Dm(T\to T_c) \simeq \frac{R_D \kB T}{6 \pi \bar\eta \xi} \simeq D_{m,0} \, \tau^{\nu x_D},
\end{equation}

where $R_D$ is a universal dimensionless number which will be discussed in Sec. IV [see \cref{dynunivratio} below]; the 
asymptotic equality on the right defines the critical amplitude $D_{m,0}$ with its dimensionless form $D_{m,0}^* := D_{m,0} \, t_0 \sigma^{-2}$. 
Note that the critical divergences of $\bar\eta$ [\cref{dynamics1,dynamic_relations} and $\xi$ imply the power-law 
singularity of $\Delta \Dm$ [\cref{DeltaDAB}] and the scaling relation $x_D = 1 + x_\eta$ [see Eqs. (1), (6), and (7)]. 
It was demonstrated before \cite{das2006jcp, das2006prl, roy2011} that the background contribution must be taken into account for a 
proper description of the simulation data. Anticipating that also the background term is proportional to temperature~\cite{das2007jcp}, 
$D_b(T) = {\mathscr L}_b(T) / \chi(T) \simeq {\mathscr L}_{\text{b},0} \kB T / \chi(T)$, suggests that the ratio 
$\mathscr L(T) / \kB T$ is described by the asymptotic law
\begin{equation}\label{onscrit2}
  \frac{\mathscr L(T)}{\kB T} \simeq {\mathscr L}_0 \tau^{-\nu x_\lambda}+{\mathscr L}_{\text{b},0} \,, \quad T \to T_c \,,
\end{equation}
with the exponent combination
\begin{equation}\label{ons5}
  \nu x_\lambda = \nu (1-\eta - x_\eta) \approx 0.567 \,,
\end{equation}
where we have used \cref{onsdef,statics1}. The connection to the amplitude of the interdiffusion constant is provided by
\begin{equation} \label{eq:DAB_amplitude}
  D_{m,0} = \mathscr{L}_0 \, \kB T_c / \chi_0 \,.
\end{equation}

We have computed $\mathscr L$ for five binary liquid mixtures for a wide range of temperatures, $1.01 T_c \leq T \lesssim 2 T_c$ 
(see \cref{fig:ons_different_density}; the data for model~II are not shown). For the three binary liquid mixtures belonging to 
model~I and within the investigated range of temperatures, ${\mathscr L}/(\kB T)$ increases by factors between 4.3 and 7.5 upon 
approaching $T_c$. This indicates the onset of the expected divergence [\cref{onscrit2}]. The remaining task is to determine 
the values of the critical amplitude $\mathscr{L}_0$ and the background contribution $\mathscr{L}_{\text{b},0}$ for each mixture 
such that \cref{onscrit2} describes the data. Here, an automated fitting routine is not suitable due to the asymptotic nature of 
power laws. Instead, the value for $\mathscr{L}_{\text{b},0}$ has been adjusted first, such that plotting 
$\Delta\mathscr{L}(T) / (\kB T) := \mathscr{L}(T) / (\kB T) - \mathscr{L}_{\text{b},0}$ as function of $\tau$ on double-logarithmic 
scales renders the data to follow straight lines of slope $-\nu x_\lambda$ for intermediate temperatures 
$0.1 \lesssim \tau \lesssim 0.5$ (\cref{fig:ons_quantification}). Indeed, subsequently for all investigated mixtures, the critical 
singularity $\Delta \mathscr{L}(T) / \kB T \sim \tau^{-\nu x_\lambda}$ [\cref{onscrit2}] can be identified in the data, which 
allows us to infer the critical amplitudes $\mathscr{L}_0$ (\cref{tab:results}).

However, for small $\tau \lesssim 0.1$, the data for $\Delta\mathscr{L}^*/T^*$ systematically deviate from the asymptotic power 
law. This is expected due to the emergence of finite-size corrections close to $T_c$~\cite{das2006prl, das2006jcp, roy2011}, 
which are significant despite the large simulation boxes used ($L/\xi \gtrsim 7$). We find that $\mathscr{L}_0$ increases by a 
factor of $\approx 6$ upon decreasing the number density $\rho$ of the fluid. On the other hand, the background contribution 
$\mathscr{L}_{\text{b},0}$ turns out to be almost insensitive to changes in the density so that the background term in 
\cref{onscrit2} becomes less relevant for smaller~$\rho$.

\subsubsection{Shear viscosity}

Another transport quantity of central interest is the shear viscosity $\bar\eta$ (not to be confused with the critical exponent 
$\eta$ of the structure factor). Due to critical slowing down, $\bar\eta(T)$ is expected to diverge at $T_c$. We have computed 
this quantity using both the Green--Kubo and the Einstein--Helfand formulae, involving the stress tensor as the generalized current. 
The Green--Kubo formula reads \cite{hansen2008, alder1970}
\begin{equation}\label{shearGK}
\bar\eta=\frac{\rho}{3 \kB T} \int_0^\infty \! \mleft[ C_{xy}(t) + C_{yz}(t) + C_{xz}(t)\mright] \diff t
\end{equation}
and is based on the autocorrelators $C_{ij}(t)$ of the off-diagonal elements of the stress tensor $\Pi_{ij}$ [\cref{stresstensor}]:
\begin{align}\label{shearGK2}
C_{ij}(t)&=\frac{1}{N} \expect{ \Pi_{ij}(t)\, \Pi_{ij}(0) }.
\end{align}
The autocorrelators $C_{ij}(t)$ are normalized by $N$ in order to render a finite value of $C_{ij}(t)$ in the thermodynamic limit.

Starting with the Helfand moments \cite{alder1970, viscardy2003}
\begin{equation}\label{helfand1}
\delta G_{ij}^2(t)=\frac{1}{N} \expect{\mleft(\int_0^t \! \Pi_{ij}(t') \,\diff t'\mright)^2} ,
\end{equation}
we have computed $\bar\eta$ alternatively by means of the Einstein--Helfand formula \cite{alder1970, viscardy2003}:
\begin{equation}\label{helfand2}
\bar\eta = \lim_{t \to \infty} \frac {\rho}{6\kB T} \frac {\diff}{\diff t}
  \mleft[\delta G^2_{xy}(t) + \delta G^2_{yz}(t) + \delta G^2_{xz}(t) \mright] \,.
\end{equation}
The expressions in \cref{shearGK2,helfand2} explicitly include averages over the different Cartesian directions due to isotropy of 
the mixed phase. We have checked that both routes yield the same values of $\bar\eta$, with the Einstein--Helfand formula generating 
smaller error bars.

\begin{figure}
\centering
\includegraphics*[width=0.4\textwidth]{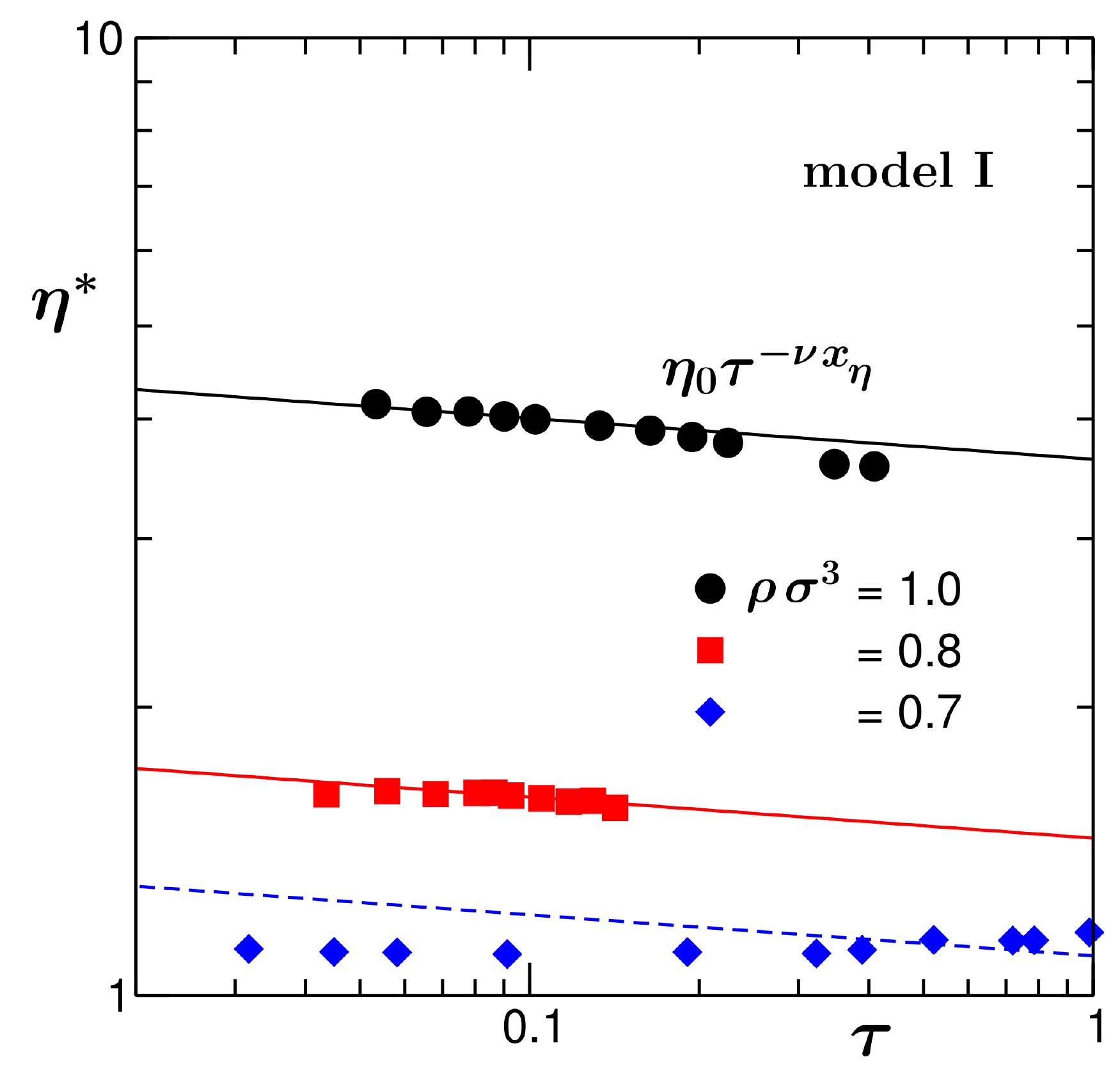}
\caption{Dimensionless shear viscosity $\bar\eta^*=\bar\eta \sigma^3 / {\varepsilon t_0}$ as a function of $\tau$, for model~I and 
three number densities. The chosen system sizes are $L=47\sigma$, $42\sigma$, and $50\sigma$ for $\rho\sigma^3=1.0$, $0.8$, and $0.7$, 
respectively. The straight lines indicate the asymptotic critical exponent $\nu x_\eta \approx 0.043$; solid lines are fits to 
the data. For $\rho\sigma^3=0.7$, the large error bars have precluded a fit of the amplitude $\eta_0$; instead, $\eta_0$ has been 
estimated from \cref{dynunivratio} using $R_D=1.0$ (dashed line). Relative errors in $\bar\eta$ vary between $2\%$ and $8\%$. 
The resulting values of $\eta_0$ are reported in Table I.} 
\label{fig:shear_quantification}
\end{figure}

The thermal singularity of $\bar\eta$ in model~$H'$ is the same as in model~$H$ and reads \cite{folk2006, bhattacharjee1982}
\begin{equation}\label{shear4}
\bar\eta \simeq \eta_0 \tau^{-\nu x_\eta} \,, \qquad \nu x_\eta \approx 0.043 \,,
\end{equation}
which can be expressed as $\bar\eta \simeq \eta_0 \xi_0^{-x_\eta} \xi^{x_\eta}$ with $\xi \simeq \xi_0 \tau^{-\nu}$ [compare 
\cref{dynamics1}]. \Cref{fig:shear_quantification} shows the shear viscosity $\bar\eta(\tau)$ for three number densities $\rho$ on 
double-logarithmic scales. The observed increase of $\bar\eta$ by a factor of $\approx 3.3$ as $\rho\sigma^3$ is varied from 0.7 to 
1.0 supports the intuitive picture that transport is slower in denser fluids. In order to facilitate the direct determination of 
$\eta_0$, instead of performing a finite-size scaling analysis \cite{roy2014}, we have considered particularly large system sizes 
(see the caption of \cref{fig:shear_quantification}). By fixing the critical exponent to $\nu x_\eta=0.043$, we have obtained 
the amplitude $\eta_0$ by fits of \cref{shear4} to the data in the temperature range that is unaffected by finite-size effects; 
the results are listed in \cref{tab:results}. The data for $\bar\eta$ at $\rho\sigma^3 = 1.0$ and $0.8$ are compatible with the 
critical power law (see solid lines in \cref{fig:shear_quantification}); the divergence, however, is hardly inferred from the 
figure due to the tiny value of the exponent $\nu x_\eta$, albeit the present error bars for $\bar\eta$ are much smaller compared 
to those reported in the literature. For $\rho\sigma^3=1.0$, due to corrections the data for $\bar\eta$ fall short of the asymptotic 
line for $\tau>0.2$. For $\rho\sigma^3=0.7$, we refrain from providing a value for $\eta_0$ because for this low density the 
determination of $\eta_0$ requires enormous statistical averaging, which we have not yet achieved. Yet, from the value $R_D=1.0$ 
of the universal amplitude ratio [\cref{dynunivratio} below] one finds $\eta_0 \simeq 1.1$. The dashed line in 
\cref{fig:shear_quantification} corresponds to this predicted value.

Actually, as in the case of the Onsager coefficient $\mathscr L$, \cref{shear4} has also to be augmented by an analytic background 
contribution $\eta_b$. For the shear viscosity, this background term has been argued to be of multiplicative character 
\cite{ohta1977}, i.e., the universal amplitude $\eta_0$ is proportional to the background viscosity and takes the form 
$\eta_0=\eta_b(q_0 \xi_0)^{x_\eta}$ with a certain (necessarily system-specific) wavenumber $q_0$ \cite{burstyn1983,das2007jcp}. 
Thus, in contrast to the case of the Onsager coefficient, the analysis of the critical divergence of the shear viscosity is not 
hampered by the presence of an analytic background.

\section{Universal amplitude ratios}
\label{sec:discussion}

Generically, critical amplitudes are non-universal and depend on microscopic details of the systems. However, certain ratios of 
critical amplitudes are known to be universal. One such ratio for static quantities is \cite{privman1991, pelissetto2002, jacobs1986}
\begin{equation}\label{twoscaleuniv}
R_\xi^+ R_c^{-1/d}=\xi_0^+ \mleft(\frac{\varphi_0^2}{\kB T_c \, \chi_0^+}\mright)^{1/d},
\end{equation}
as predicted by the hypothesis of two-scale factor universality. Here, the superscript ``$+$" emphasizes that (apart from 
$\varphi_0$) the amplitudes correspond to $T> T_c$. For binary liquid mixtures belonging to the 3d Ising universality class, 
the value of $R_\xi^+ R_c^{-1/d}$, as estimated theoretically and experimentally, lies within the ranges $[0.68, 0.70]$ and 
$[0.67, 0.72]$, respectively \cite{pelissetto2002}. 

The so-called Kawasaki amplitude $R_D=6 \pi \bar\eta \xi/(\kB T \Delta \Dm)$ [\cref{DeltaDAB}] is a universal amplitude ratio 
involving transport coefficients, i.e., the critical enhancement $\Delta\Dm$ of the mutual diffusivity [\cref{DeltaDAB}]. 
Inserting the asymptotic singular behaviors of $\xi$, $\chi$, and $\bar\eta$ [\cref{criticallawsus,shear4}] as well as 
$\Delta\Dm \simeq {\mathscr L}_0 \tau^{-\nu x_\lambda} / \chi_0$ [see \cref{onsdef,onscrit2}], the temperature dependence drops out and one finds
\begin{equation}\label{dynunivratio}
  R_D=\frac{6\pi\eta_0 \xi_0 {\mathscr L}_0}{\chi_0} \,.
\end{equation}
This combination of non-universal static and dynamic critical amplitudes has been shown to be a universal number~\cite{sengers1985}. 
Theoretical calculations based on dynamic RGT predict $R_D \approx 1.07$~\cite{paladin1982}, while MCT provides 
$R_D\approx 1.03$ \cite{burstyn1983}; experimental data yield $R_D=1.01\pm 0.04$~\cite{burstyn1983,sengers1985}.

\begin{figure}
\centering
\includegraphics*[width=0.4\textwidth]{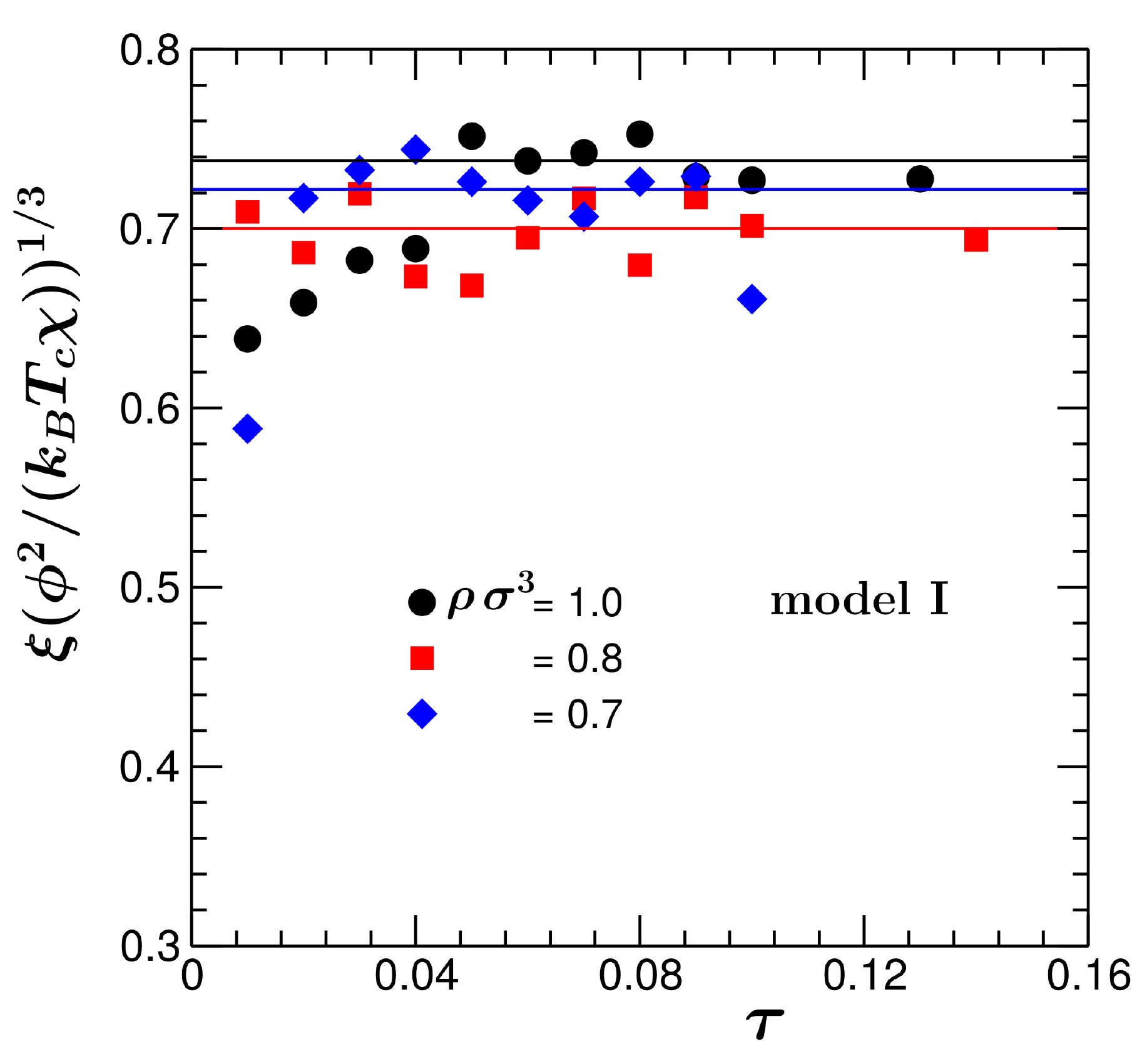}
\caption{Temperature dependence of the dimensionless and supposedly universal quantity $\xi (\phi^2/ (\kB T_c \chi))^{1/3}$ 
within model I for three number densities. Symbols correspond to simulation data and solid lines represent averages of the data points. 
}
\label{fig:amplituderatio}
\end{figure}

\begin{figure}
\centering
\includegraphics*[width=0.4\textwidth]{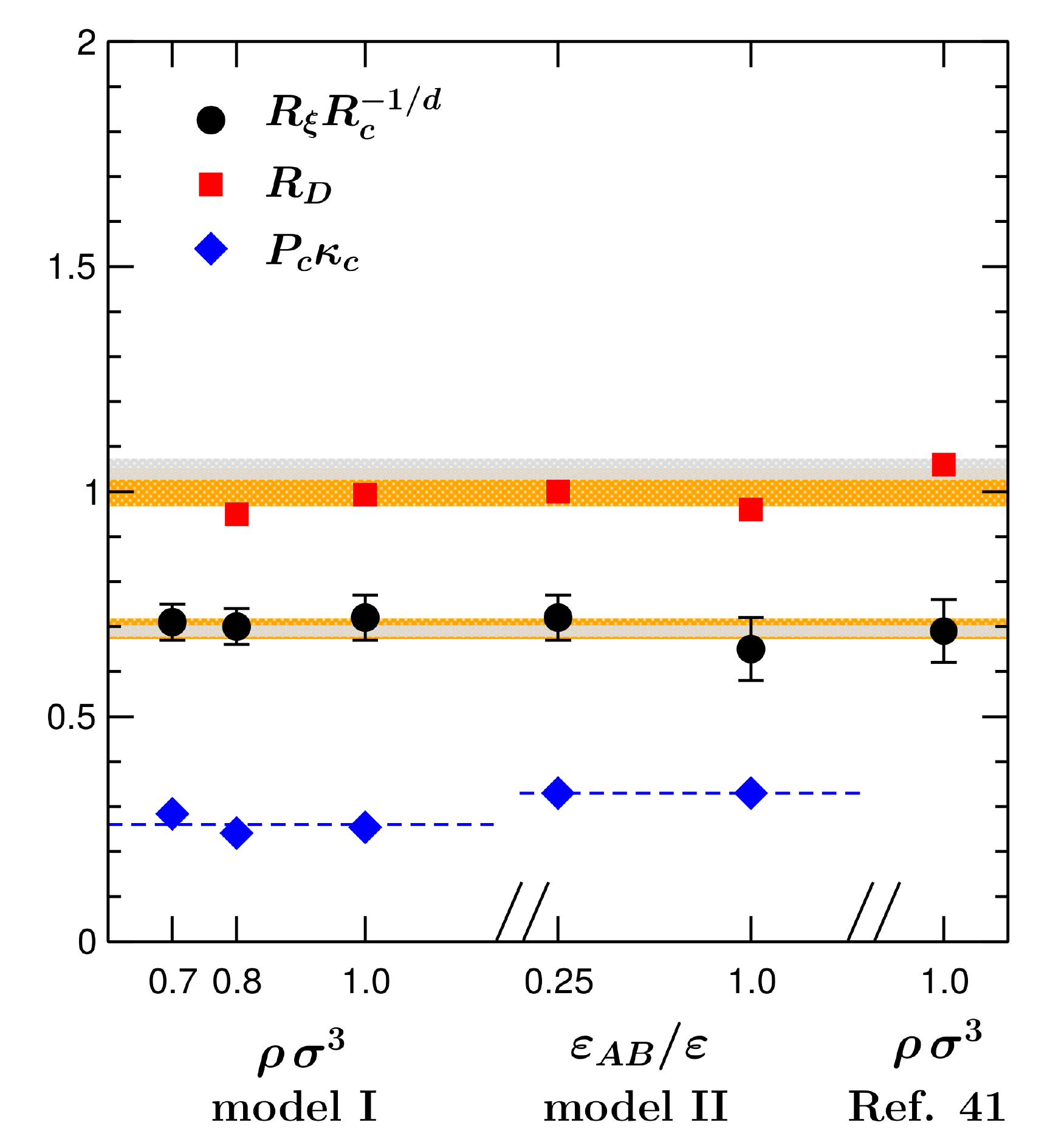}
\caption{The universal static amplitude ratio $R_{\xi}R_c^{-1/d}$, the dynamic universal amplitude ratio $R_D$, and the 
dimensionless quantity $P_c \kappa_c$ at criticality for various binary liquid mixtures within models I and II. The grey and 
orange regions represent the ranges of the theoretical and experimental predictions, respectively. The dashed lines correspond 
to the averages for $P_c \kappa_c$ within models I and II, respectively. The relative error bars of $R_D$ are about $20\%$.}
\label{fig:universalratio}
\end{figure}

A calculation of the amplitude ratio $R_\xi^+ R_c^{-1/d}$ in \cref{twoscaleuniv} combines the uncertainties in the separately determined 
amplitudes $\varphi_0$, $\xi_0$, and $\chi_0$. \cref{tab:results} lists these values. Equivalently, the universal ratio is 
given directly as the limit $\tau \searrow 0$ of the combination $Z(\tau) := \xi(\tau) [\phi(\tau)^2/ (\kB T_c \chi(\tau))]^{1/3}$. 
However, the omnipresent finite-size corrections prohibit us from taking the limit rigorously. Yet, one can expect to find a 
temperature range close to $T_c$ in which all quantities $\phi$, $\xi$, and $\chi$ follow their asymptotic critical laws. This 
implies that in this temperature range $Z(\tau)$ displays a plateau at the value of $R_{\xi}^+ R_c^{-1/d}$. \Cref{fig:amplituderatio} 
provides a test of this approach for the three mixtures within model~I. Indeed, a plateau may be inferred for each data set 
after averaging out the scatter of the data points. The estimates of $R_{\xi}^+ R_c^{-1/d}$ obtained this way 
($0.738\pm 0.016,~ 0.70 \pm 0.03, ~ 0.722\pm 0.023$ for $\rho\sigma^3=1.0,~0.8,~0.7$, respectively, within model~I) match well 
with those obtained from \cref{twoscaleuniv} by inserting the critical amplitudes, but exhibit slightly smaller errors. The results 
for the 5 binary mixtures studied here as well as the results of Ref.~\cite{das2006jcp} corroborate that 
$R_{\xi}^+ R_c^{-1/d}$ is a universal number with a value of $0.70\pm0.01$ (\cref{fig:universalratio}). Our estimate for 
$R_{\xi}^+ R_c^{-1/d}$ is in nice agreement with previous values for this universal ratio obtained from theory and experiments 
[see text below \cref{twoscaleuniv}].

Concerning the dynamic amplitude ratio $R_D$, we report results for model~I only for $\rho\sigma^3=1.0$ and $0.8$ because it is 
difficult to resolve the critical behavior of the viscosity at low densities. A similar analysis as above in terms of 
$Y(\tau)=6 \pi\eta\xi {\Delta \mathscr L}/{k_B T \chi}$ has turned out to be inconclusive, in that no plateau in $Y(\tau)$ has emerged. 
We attribute this to the fact that the critical range of temperatures (free from both asymptotic and finite-size corrections) for 
the Onsager coefficient is located at higher temperatures than for the other quantities entering \cref{dynunivratio} 
(see also Figs. 6, 7, 9, and 10). Therefore, \cref{tab:results} lists the values for $R_D$ as obtained from \cref{dynunivratio}. 
Despite significant error bars of about 20\%, the estimates coincide surprisingly well with the expectation $R_D\approx 1.0$ (\cref{fig:universalratio})

Finally, we note that the dimensionless product $P_c \kappa_c$ of pressure and compressibility at the demixing transition appears 
to stay almost constant at $0.26 \pm 0.02$ within model~I (insensitive to the density $\rho$) and at ca. 0.33 for model~II 
(insensitive to the strength $\epsilon_\text{AB}$ of repulsion). This is remarkable because $P_c$ and $\kappa_c$ separately 
vary across these ranges by almost an order of magnitude. However, here we point out that the product $P_c \kappa_c$ is not 
related to the order parameter field, which is the concentration, but to the number density field, which in model~$H'$ serves 
as a secondary conserved field \cite{folk2006}. Thus, there is no theoretical basis to consider $P_c \kappa_c$ as a universal 
number; indeed the values of $P_c \kappa_c$ are different for models I and~II.

\section{Summary and conclusions}
\label{sec:conclusion}

We have computationally investigated the static and dynamic properties of five symmetric binary liquid mixtures close to their 
continuous demixing transitions. To this end, we have employed a combination of Monte Carlo simulations in the semi-grand 
canonical ensemble and molecular dynamics (MD) simulations. While the former is suited best to determine the phase diagram, 
only the latter obeys the conservation laws of actual liquid mixtures and thus properly captures the critical dynamics associated 
with model~$H'$. Previous computational studies of the critical behavior of such mixtures have been based on small system 
sizes in conjunction with suitable finite-size scaling analyses. A massively parallel implementation of the MD simulations using 
GPUs made it possible to explore much larger system sizes than before, which has allowed us to determine the critical amplitudes directly.

The chosen mixtures represent a wide range of critical temperatures $T_c$, number densities $\rho$, and isothermal compressibilities~$\kappa_T$. 
For all mixtures considered, the particles interact via truncated Lennard-Jones potentials. The interaction potential $u_\text{AB}(r)$ 
for pairs of unlike particles has been chosen to either include the usual attractive part or to be purely repulsive, which we 
refer to as models~I and II, respectively. For the fluids in model~I, the density $\rho$ has been varied, while within model~II 
the strength $\varepsilon_\text{AB}$ of the repulsion between unlike species has been varied. All results of the data analysis 
have been compiled in \cref{tab:results}. The main findings of our work are the following:

\emph{(i)} For each fluid, we have calculated the phase diagram in the temperature--composition plane, from which the corresponding 
critical temperatures $T_c$ have been extracted by using the critical scaling of the order parameter (\cref{fig:phasedia}) and Binder's 
intersection method (\cref{fig:bincum}). We have found that the values of $T_c$ within model~II are a factor of ca. 2 higher 
than for otherwise comparable fluids in model~I. Within model~II, reinforcing the repulsion $\varepsilon_{AB}$ leads to a drastic increase of~$T_c$. 
Further, at the demixing transition, we have computed the pressure $P_c$ and the isothermal compressibility $\kappa_c$, which exhibit 
a large variability across all mixtures, covering almost one order of magnitude.

\emph{(ii)} The loci of the liquid--liquid critical points $T_c(\rho)$, also referred to as $\lambda$-line, has been calculated 
within model~I. This curve $T_c(\rho)$ is non-monotonic, indicating a re-entrance phenomenon upon varying the density along an 
isotherm (\cref{fig:lambdaline}). In this context, for model~I we have also investigated the potential occurrence of a critical 
end point or a tri-critical point at which the $\lambda$-line meets the liquid--vapor critical point. Our results for the isothermal 
compressibility (\cref{fig:snn_allrho}) indicate at that this is not the case. This issue calls for additional future investigations.

\emph{(iii)} The structural properties of the mixtures have been analyzed in terms of the static structure factors $S_{cc}(k)$ 
and $S_{\rho \rho}(k)$ of the composition and density fields, respectively (\cref{fig:scc}). As expected, long-wavelength fluctuations 
of the composition become dominant near the demixing transition: for small wave numbers, $S_{cc}(k\to 0)$ increases sharply as $T_c$ 
is approached, at which the critical power law $S_{cc}(k) \sim k^{-2+\eta}$ is observed over one decade in~$k$ which is 
facilitated by the large system sizes chosen. To the contrary, $S_{\rho\rho}(k)$, which is probing density fluctuations, is almost 
insensitive to temperature changes in the range $T_c \leq T < 1.6 T_c$; in particular, it does not display any critical enhancement at small~$k$.

\emph{(iv)} From $S_{cc}(k)$ we have determined the correlation length $\xi$ and the order parameter susceptibility $\chi$. For both quantities, 
the scaling with the corresponding critical Ising exponents is confirmed (\cref{fig:sus,fig:cor}), allowing us to extract the 
non-universal critical amplitudes $\xi_0$ and $\chi_0$. We have found that $\chi_0$ decreases upon increasing density (model~I), 
which we attribute to an energetically penalized particle rearrangement at denser packing. The values of $\chi_0$ are less 
sensitive to changes in the strength of the AB repulsion (model~II). The correlation length $\xi$ is limited by the finite system size. 
Nonetheless we have been able to achieve values of up to $\xi \approx 10\sigma$ in our simulations. Across all five binary 
fluid mixtures, its amplitude varies only mildly around $\xi_0 \approx 0.5\sigma$.

\emph{(v)} The critical transport behavior has been studied in terms of the Onsager coefficient $\mathscr L=\chi D_m$ and the shear 
viscosity $\bar\eta$, the former also determining the interdiffusion constant $D_m$. Within model~I, the Onsager coefficient 
and thus its critical amplitude ${\mathscr L}_0$ increase by a factor of~6 upon varying the density from $\rho\sigma^3=1$ to $0.7$ 
(\cref{fig:ons_quantification,fig:ons_different_density}); concomitantly, the shear viscosity $\bar\eta$ decreases by a 
factor of~3 (\cref{fig:shear_quantification}). This trend is in line with our notion that mass diffusion is faster in a less dense 
fluid; it also has direct consequences for the computational efficiency of a model. The asymptotic critical enhancement of 
$\mathscr L$ is obscured, first, by the non-universal analytic background contribution away from $T_c$ and, second, by 
finite-size corrections close to $T_c$, which are still significant despite the large simulation boxes we used. These issues have 
prevented us to obtain accurate estimates of $\mathscr{L}_0$. Furthermore, the critical behavior of 
$\bar\eta \sim |T-T_c|^{-\nu x_\eta}, \: \nu x_\eta \approx 0.043$, is difficult to assess reliably due to the smallness of the 
critical exponent. We have obtained the critical amplitude for 3 out of the 5 mixtures (within model~I for $\rho\sigma^3=1,\: 0.8$ 
and within model~II for $\epsilon_{AB}=0.25 \varepsilon$).

\emph{(vi)} Finally, we have computed two universal amplitude ratios, involving several static and dynamic non-universal critical 
amplitudes (all above $T_c$). One such ratio of static quantities is $R_\xi^+ R_c^{-1/3}$ [\cref{twoscaleuniv}], the other ratio 
$R_D$ is a combination of both static and dynamic amplitudes [\cref{dynunivratio}]. For both ratios, quantitative predictions 
are available for the universality classes of the models~$H~\text{and}~H'$ based on mode-coupling and dynamic renormalization 
group theories, which are supported by experimental data. Across all 5 mixtures studied and including the results of 
Ref.~\cite{das2006jcp}, the simulation results for the static ratio $R_\xi^+ R_c^{-1/3}$ yield a universal value $0.70 \pm 0.01$, 
in agreement with theoretical predictions (\cref{fig:universalratio}). Our results for the dynamic ratio $R_D$ are compatible with 
theoretical and experimental estimates, but they are subject to large uncertainties given the difficulties in determining the 
dynamic critical amplitudes $\mathscr{L}_0$ and $\eta_0$. A notable finding is that the dimensionless product $P_c \kappa_c$ of 
pressure and compressibility is remarkably constant along the $\lambda$-line in model~I and with respect to variations of the 
strength~$\varepsilon_\text{AB}$ of repulsion in model~II (\cref{fig:universalratio}).

The present study reports the first comprehensive analysis of the density dependence of the critical amplitudes. The knowledge 
of these amplitudes for a given simulation model facilitates the calibration of the model to a given physical binary liquid mixture. 
As an example we refer to the well-characterized water--lutidine mixture \cite{mirzaev2006} and model~I for $\rho \sigma^3 = 0.7$: 
first, the measured correlation length amplitude $\xi_0 \approx \SI{0.2}{nm}$ implies the length scale $\sigma \approx \SI{0.4}{nm}$ for 
all of the presented simulations. Second, the relaxation rate amplitude $\Gamma_0 \approx \SI{25e9}{s^{-1}}$ yields the critical 
amplitude of the interdiffusion constant $D_{m,0} = \Gamma_0 \xi_0^2 / 2 \approx \SI{5e-10}{m^2 s^{-1}}$, which has to be compared 
with the simulation result $D_{m,0} \approx 0.09 \sigma^2/t_0$, fixing the time scale $t_0 \approx \SI{30}{ps}$. The energy scale 
$\varepsilon$ is set by the critical temperature, $T_c \approx \SI{307}{K}$ in the experiments and $\kB T_c \approx 1.51 \varepsilon$ 
in the simulations, and thus $\varepsilon / \kB \approx \SI{203}{K}$ or $\varepsilon \approx \SI{2.8e-21}{J}$. Accordingly, the 
universal amplitude ratios fix the critical amplitudes for a number of related physical quantities such as the composition susceptibility 
$\chi_0$, the viscosity $\eta_0$, or the surface tension \cite{das2011}. The coarse-grained simulation models 
(replacing an organic molecule and water by symmetric Lennard-Jones spheres), however, come at the price that physical quantities not 
linked to the critical singularities may not be captured correctly. For instance, along these lines, the pressure at criticality is 
expected to be $P_c \approx \SI{110}{MPa}$, which is 3 orders of magnitude larger than the ambient pressure, while the compressibility 
$\kappa_c \approx \SI{25e-10}{Pa^{-1}}$ is too high as well. Increasing the density $\rho$ will reduce the compressibility, but simultaneously 
increase the pressure and also slow down the overall dynamics (which is computationally expensive). Quantitative agreement with 
actual binary liquid mixtures can be achieved with force-field-based simulation models, see \cite{guevara2016} for a recent study. 
Nevertheless, the comparably simple models discussed here can correctly describe the physical behavior at long wave length thanks 
to the universality of the demixing transition. The presented compilation of results may serve as a guide to find the simulation model 
that is best suited to address a specific phenomenon.

This study is supposed to stimulate further computational investigations concerning critical transport in fluids. Specifically, so far 
there are no dedicated computations of dynamic critical amplitudes \textit{below} $T_c$ and also none for liquid--vapor transitions, 
neither above nor below $T_c$. A quantitatively reliable determination of the ratio $\eta_0/\eta_b$ is also of significant importance, 
in particular in view of the difficulties associated with obtaining an accurate value of this ratio from experiments.

\begin{acknowledgments}
We acknowledge the use of the supercomputer \emph{Hydra} of the Max Planck Computing and Data Facility Garching for producing most of the MD data. 
\end{acknowledgments}

\bibliography{critical_fluids}

\begin{thebibliography}{83}%
\makeatletter
\providecommand \@ifxundefined [1]{%
 \@ifx{#1\undefined}
}%
\providecommand \@ifnum [1]{%
 \ifnum #1\expandafter \@firstoftwo
 \else \expandafter \@secondoftwo
 \fi
}%
\providecommand \@ifx [1]{%
 \ifx #1\expandafter \@firstoftwo
 \else \expandafter \@secondoftwo
 \fi
}%
\providecommand \natexlab [1]{#1}%
\providecommand \enquote  [1]{``#1''}%
\providecommand \bibnamefont  [1]{#1}%
\providecommand \bibfnamefont [1]{#1}%
\providecommand \citenamefont [1]{#1}%
\providecommand \href@noop [0]{\@secondoftwo}%
\providecommand \href [0]{\begingroup \@sanitize@url \@href}%
\providecommand \@href[1]{\@@startlink{#1}\@@href}%
\providecommand \@@href[1]{\endgroup#1\@@endlink}%
\providecommand \@sanitize@url [0]{\catcode `\\12\catcode `\$12\catcode
  `\&12\catcode `\#12\catcode `\^12\catcode `\_12\catcode `\%12\relax}%
\providecommand \@@startlink[1]{}%
\providecommand \@@endlink[0]{}%
\providecommand \url  [0]{\begingroup\@sanitize@url \@url }%
\providecommand \@url [1]{\endgroup\@href {#1}{\urlprefix }}%
\providecommand \urlprefix  [0]{URL }%
\providecommand \Eprint [0]{\href }%
\providecommand \doibase [0]{http://dx.doi.org/}%
\providecommand \selectlanguage [0]{\@gobble}%
\providecommand \bibinfo  [0]{\@secondoftwo}%
\providecommand \bibfield  [0]{\@secondoftwo}%
\providecommand \translation [1]{[#1]}%
\providecommand \BibitemOpen [0]{}%
\providecommand \bibitemStop [0]{}%
\providecommand \bibitemNoStop [0]{.\EOS\space}%
\providecommand \EOS [0]{\spacefactor3000\relax}%
\providecommand \BibitemShut  [1]{\csname bibitem#1\endcsname}%
\let\auto@bib@innerbib\@empty
\bibitem [{\citenamefont {Fisher}(1967)}]{fisher1967}%
  \BibitemOpen
  \bibfield  {author} {\bibinfo {author} {\bibfnamefont {M.~E.}\ \bibnamefont
  {Fisher}},\ }\href
  {http://www.scopus.com/inward/record.url?eid=2-s2.0-0346365713&partnerID=40&md5=162291a25cc8353b0e8fc0c38e7f6b79}
  {\bibfield  {journal} {\bibinfo  {journal} {Rep. Prog. Phys.}\ }\textbf
  {\bibinfo {volume} {30}},\ \bibinfo {pages} {615} (\bibinfo {year}
  {1967})}\BibitemShut {NoStop}%
\bibitem [{\citenamefont {Stanley}(1971)}]{stanley1971}%
  \BibitemOpen
  \bibfield  {author} {\bibinfo {author} {\bibfnamefont {H.}~\bibnamefont
  {Stanley}},\ }\href@noop {} {\emph {\bibinfo {title} {Introduction to Phase
  Transitions and Critical Phenomena}}}\ (\bibinfo  {publisher} {Oxford
  University Press},\ \bibinfo {address} {Oxford},\ \bibinfo {year}
  {1971})\BibitemShut {NoStop}%
\bibitem [{\citenamefont {Hohenberg}\ and\ \citenamefont
  {Halperin}(1977)}]{hohenberg1977}%
  \BibitemOpen
  \bibfield  {author} {\bibinfo {author} {\bibfnamefont {P.~C.}\ \bibnamefont
  {Hohenberg}}\ and\ \bibinfo {author} {\bibfnamefont {B.~I.}\ \bibnamefont
  {Halperin}},\ }\href {\doibase 10.1103/RevModPhys.49.435} {\bibfield
  {journal} {\bibinfo  {journal} {Rev. Mod. Phys.}\ }\textbf {\bibinfo {volume}
  {49}},\ \bibinfo {pages} {435} (\bibinfo {year} {1977})}\BibitemShut
  {NoStop}%
\bibitem [{\citenamefont {Privman}\ \emph {et~al.}(1991)\citenamefont
  {Privman}, \citenamefont {Hohenberg},\ and\ \citenamefont
  {Aharony}}]{privman1991}%
  \BibitemOpen
  \bibfield  {author} {\bibinfo {author} {\bibfnamefont {V.}~\bibnamefont
  {Privman}}, \bibinfo {author} {\bibfnamefont {P.}~\bibnamefont {Hohenberg}},
  \ and\ \bibinfo {author} {\bibfnamefont {A.}~\bibnamefont {Aharony}},\ }in\
  \href {http://people.clarkson.edu/~vprivman/77.pdf} {\emph {\bibinfo
  {booktitle} {Phase Transitions and Critical Phenomena}}},\ Vol.~\bibinfo
  {volume} {14},\ \bibinfo {editor} {edited by\ \bibinfo {editor}
  {\bibfnamefont {C.}~\bibnamefont {Domb}}\ and\ \bibinfo {editor}
  {\bibfnamefont {J.}~\bibnamefont {Lebowitz}}}\ (\bibinfo  {publisher}
  {Academic},\ \bibinfo {address} {New York},\ \bibinfo {year} {1991})\
  Chap.~\bibinfo {chapter} {1}, pp.\ \bibinfo {pages} {1--134}\BibitemShut
  {NoStop}%
\bibitem [{\citenamefont {Pelissetto}\ and\ \citenamefont
  {Vicardi}(2002)}]{pelissetto2002}%
  \BibitemOpen
  \bibfield  {author} {\bibinfo {author} {\bibfnamefont {A.}~\bibnamefont
  {Pelissetto}}\ and\ \bibinfo {author} {\bibfnamefont {E.}~\bibnamefont
  {Vicardi}},\ }\href {\doibase 10.1016/S0370-1573(02)00219-3} {\bibfield
  {journal} {\bibinfo  {journal} {Phys. Rep.}\ }\textbf {\bibinfo {volume}
  {368}},\ \bibinfo {pages} {549} (\bibinfo {year} {2002})}\BibitemShut
  {NoStop}%
\bibitem [{\citenamefont {Folk}\ and\ \citenamefont {Moser}(2006)}]{folk2006}%
  \BibitemOpen
  \bibfield  {author} {\bibinfo {author} {\bibfnamefont {R.}~\bibnamefont
  {Folk}}\ and\ \bibinfo {author} {\bibfnamefont {G.}~\bibnamefont {Moser}},\
  }\href {http://stacks.iop.org/0305-4470/39/i=24/a=R01} {\bibfield  {journal}
  {\bibinfo  {journal} {J. Phys. A}\ }\textbf {\bibinfo {volume} {39}},\
  \bibinfo {pages} {R207} (\bibinfo {year} {2006})}\BibitemShut {NoStop}%
\bibitem [{\citenamefont {Fisher}(1998)}]{fisher1998}%
  \BibitemOpen
  \bibfield  {author} {\bibinfo {author} {\bibfnamefont {M.~E.}\ \bibnamefont
  {Fisher}},\ }\href {\doibase 10.1103/RevModPhys.70.653} {\bibfield  {journal}
  {\bibinfo  {journal} {Rev. Mod. Phys.}\ }\textbf {\bibinfo {volume} {70}},\
  \bibinfo {pages} {653} (\bibinfo {year} {1998})}\BibitemShut {NoStop}%
\bibitem [{\citenamefont {Heller}(1967)}]{heller1967}%
  \BibitemOpen
  \bibfield  {author} {\bibinfo {author} {\bibfnamefont {P.}~\bibnamefont
  {Heller}},\ }\href@noop {} {\bibfield  {journal} {\bibinfo  {journal} {Rep.
  Prog. Phys.}\ }\textbf {\bibinfo {volume} {30}},\ \bibinfo {pages} {731}
  (\bibinfo {year} {1967})}\BibitemShut {NoStop}%
\bibitem [{\citenamefont {Sengers}\ and\ \citenamefont
  {Shanks}(2009)}]{sengers2009}%
  \BibitemOpen
  \bibfield  {author} {\bibinfo {author} {\bibfnamefont {J.~V.}\ \bibnamefont
  {Sengers}}\ and\ \bibinfo {author} {\bibfnamefont {J.~G.}\ \bibnamefont
  {Shanks}},\ }\href {\doibase 10.1007/s10955-009-9840-z} {\bibfield  {journal}
  {\bibinfo  {journal} {J. Stat. Phys.}\ }\textbf {\bibinfo {volume} {137}},\
  \bibinfo {pages} {857} (\bibinfo {year} {2009})}\BibitemShut {NoStop}%
\bibitem [{\citenamefont {Hertlein}\ \emph {et~al.}(2008)\citenamefont
  {Hertlein}, \citenamefont {Helden}, \citenamefont {Gambassi}, \citenamefont
  {Dietrich},\ and\ \citenamefont {Bechinger}}]{hertlein2008}%
  \BibitemOpen
  \bibfield  {author} {\bibinfo {author} {\bibfnamefont {C.}~\bibnamefont
  {Hertlein}}, \bibinfo {author} {\bibfnamefont {L.}~\bibnamefont {Helden}},
  \bibinfo {author} {\bibfnamefont {A.}~\bibnamefont {Gambassi}}, \bibinfo
  {author} {\bibfnamefont {S.}~\bibnamefont {Dietrich}}, \ and\ \bibinfo
  {author} {\bibfnamefont {C.}~\bibnamefont {Bechinger}},\ }\href
  {http://dx.doi.org/10.1038/nature06443} {\bibfield  {journal} {\bibinfo
  {journal} {Nature}\ }\textbf {\bibinfo {volume} {451}},\ \bibinfo {pages}
  {172} (\bibinfo {year} {2008})}\BibitemShut {NoStop}%
\bibitem [{\citenamefont {Buttinoni}\ \emph {et~al.}(2012)\citenamefont
  {Buttinoni}, \citenamefont {Volpe}, \citenamefont {K{\"u}mmel}, \citenamefont
  {Volpe},\ and\ \citenamefont {Bechinger}}]{buttinoni2012}%
  \BibitemOpen
  \bibfield  {author} {\bibinfo {author} {\bibfnamefont {I.}~\bibnamefont
  {Buttinoni}}, \bibinfo {author} {\bibfnamefont {G.}~\bibnamefont {Volpe}},
  \bibinfo {author} {\bibfnamefont {F.}~\bibnamefont {K{\"u}mmel}}, \bibinfo
  {author} {\bibfnamefont {G.}~\bibnamefont {Volpe}}, \ and\ \bibinfo {author}
  {\bibfnamefont {C.}~\bibnamefont {Bechinger}},\ }\href
  {http://stacks.iop.org/0953-8984/24/i=28/a=284129} {\bibfield  {journal}
  {\bibinfo  {journal} {J. Phys: Condens. Matter}\ }\textbf {\bibinfo {volume}
  {24}},\ \bibinfo {pages} {284129} (\bibinfo {year} {2012})}\BibitemShut
  {NoStop}%
\bibitem [{\citenamefont {Anisimov}\ and\ \citenamefont
  {Sengers}(2000)}]{anisimov2000}%
  \BibitemOpen
  \bibfield  {author} {\bibinfo {author} {\bibfnamefont {M.}~\bibnamefont
  {Anisimov}}\ and\ \bibinfo {author} {\bibfnamefont {J.}~\bibnamefont
  {Sengers}},\ }in\ \href {\doibase 10.1016/S1874-5644(00)80022-3} {\emph
  {\bibinfo {booktitle} {Equations of State for Fluids and Fluid Mixtures}}},\
  \bibinfo {series and number} {Experimental Thermodynamics V},\ \bibinfo
  {editor} {edited by\ \bibinfo {editor} {\bibfnamefont {J.}~\bibnamefont
  {Sengers}}, \bibinfo {editor} {\bibfnamefont {R.}~\bibnamefont {Kayser}},
  \bibinfo {editor} {\bibfnamefont {C.}~\bibnamefont {Peters}}, \ and\ \bibinfo
  {editor} {\bibfnamefont {H.}~\bibnamefont {White}}}\ (\bibinfo  {publisher}
  {Elsevier},\ \bibinfo {address} {Amsterdam},\ \bibinfo {year} {2000})\
  Chap.~\bibinfo {chapter} {11}, pp.\ \bibinfo {pages} {381--434}\BibitemShut
  {NoStop}%
\bibitem [{\citenamefont {Binder}(2010)}]{binder2010}%
  \BibitemOpen
  \bibfield  {author} {\bibinfo {author} {\bibfnamefont {K.}~\bibnamefont
  {Binder}},\ }\href {\doibase 10.1080/00268976.2010.495734} {\bibfield
  {journal} {\bibinfo  {journal} {Mol. Phys.}\ }\textbf {\bibinfo {volume}
  {108}},\ \bibinfo {pages} {1797} (\bibinfo {year} {2010})}\BibitemShut
  {NoStop}%
\bibitem [{\citenamefont {Binder}(1981)}]{binder1981}%
  \BibitemOpen
  \bibfield  {author} {\bibinfo {author} {\bibfnamefont {K.}~\bibnamefont
  {Binder}},\ }\href {\doibase 10.1007/BF01293604} {\bibfield  {journal}
  {\bibinfo  {journal} {Z. Phys. B}\ }\textbf {\bibinfo {volume} {43}},\
  \bibinfo {pages} {119} (\bibinfo {year} {1981})}\BibitemShut {NoStop}%
\bibitem [{\citenamefont {Landau}\ and\ \citenamefont
  {Binder}(2009)}]{landau2009}%
  \BibitemOpen
  \bibfield  {author} {\bibinfo {author} {\bibfnamefont {D.}~\bibnamefont
  {Landau}}\ and\ \bibinfo {author} {\bibfnamefont {K.}~\bibnamefont
  {Binder}},\ }\href@noop {} {\emph {\bibinfo {title} {A Guide to Monte Carlo
  Simulations in Statistical Physics}}},\ \bibinfo {edition} {3rd}\ ed.\
  (\bibinfo  {publisher} {Cambridge University Press},\ \bibinfo {address}
  {Cambridge},\ \bibinfo {year} {2009})\BibitemShut {NoStop}%
\bibitem [{\citenamefont {Wilding}\ and\ \citenamefont
  {Nielaba}(1996)}]{wilding1996}%
  \BibitemOpen
  \bibfield  {author} {\bibinfo {author} {\bibfnamefont {N.~B.}\ \bibnamefont
  {Wilding}}\ and\ \bibinfo {author} {\bibfnamefont {P.}~\bibnamefont
  {Nielaba}},\ }\href {\doibase 10.1103/PhysRevE.53.926} {\bibfield  {journal}
  {\bibinfo  {journal} {Phys. Rev. E}\ }\textbf {\bibinfo {volume} {53}},\
  \bibinfo {pages} {926} (\bibinfo {year} {1996})}\BibitemShut {NoStop}%
\bibitem [{\citenamefont {Onuki}(1997)}]{onuki1997}%
  \BibitemOpen
  \bibfield  {author} {\bibinfo {author} {\bibfnamefont {A.}~\bibnamefont
  {Onuki}},\ }\href {\doibase 10.1103/PhysRevE.55.403} {\bibfield  {journal}
  {\bibinfo  {journal} {Phys. Rev. E}\ }\textbf {\bibinfo {volume} {55}},\
  \bibinfo {pages} {403} (\bibinfo {year} {1997})}\BibitemShut {NoStop}%
\bibitem [{\citenamefont {Bhattacharjee}\ \emph {et~al.}(2010)\citenamefont
  {Bhattacharjee}, \citenamefont {Kaatze},\ and\ \citenamefont
  {Mirzaev}}]{bhattacharjee2010}%
  \BibitemOpen
  \bibfield  {author} {\bibinfo {author} {\bibfnamefont {J.~K.}\ \bibnamefont
  {Bhattacharjee}}, \bibinfo {author} {\bibfnamefont {U.}~\bibnamefont
  {Kaatze}}, \ and\ \bibinfo {author} {\bibfnamefont {S.~Z.}\ \bibnamefont
  {Mirzaev}},\ }\href {http://stacks.iop.org/0034-4885/73/i=6/a=066601}
  {\bibfield  {journal} {\bibinfo  {journal} {Rep. Prog. Phys.}\ }\textbf
  {\bibinfo {volume} {73}},\ \bibinfo {pages} {066601} (\bibinfo {year}
  {2010})}\BibitemShut {NoStop}%
\bibitem [{\citenamefont {Furukawa}\ \emph {et~al.}(2013)\citenamefont
  {Furukawa}, \citenamefont {Gambassi}, \citenamefont {Dietrich},\ and\
  \citenamefont {Tanaka}}]{furukawa2013}%
  \BibitemOpen
  \bibfield  {author} {\bibinfo {author} {\bibfnamefont {A.}~\bibnamefont
  {Furukawa}}, \bibinfo {author} {\bibfnamefont {A.}~\bibnamefont {Gambassi}},
  \bibinfo {author} {\bibfnamefont {S.}~\bibnamefont {Dietrich}}, \ and\
  \bibinfo {author} {\bibfnamefont {H.}~\bibnamefont {Tanaka}},\ }\href
  {\doibase 10.1103/PhysRevLett.111.055701} {\bibfield  {journal} {\bibinfo
  {journal} {Phys. Rev. Lett.}\ }\textbf {\bibinfo {volume} {111}},\ \bibinfo
  {pages} {055701} (\bibinfo {year} {2013})}\BibitemShut {NoStop}%
\bibitem [{\citenamefont {Onuki}(2002)}]{onuki2002}%
  \BibitemOpen
  \bibfield  {author} {\bibinfo {author} {\bibfnamefont {A.}~\bibnamefont
  {Onuki}},\ }\href@noop {} {\emph {\bibinfo {title} {Phase Transition
  Dynamics}}}\ (\bibinfo  {publisher} {Cambridge University Press},\ \bibinfo
  {address} {Cambridge},\ \bibinfo {year} {2002})\BibitemShut {NoStop}%
\bibitem [{\citenamefont {Folk}\ and\ \citenamefont
  {Moser}(1995{\natexlab{a}})}]{folk1995}%
  \BibitemOpen
  \bibfield  {author} {\bibinfo {author} {\bibfnamefont {R.}~\bibnamefont
  {Folk}}\ and\ \bibinfo {author} {\bibfnamefont {G.}~\bibnamefont {Moser}},\
  }\href {\doibase 10.1007/BF02083546} {\bibfield  {journal} {\bibinfo
  {journal} {Int. J. Thermophys.}\ }\textbf {\bibinfo {volume} {16}},\ \bibinfo
  {pages} {1363} (\bibinfo {year} {1995}{\natexlab{a}})}\BibitemShut {NoStop}%
\bibitem [{\citenamefont {Folk}\ and\ \citenamefont
  {Moser}(1995{\natexlab{b}})}]{folk1995a}%
  \BibitemOpen
  \bibfield  {author} {\bibinfo {author} {\bibfnamefont {R.}~\bibnamefont
  {Folk}}\ and\ \bibinfo {author} {\bibfnamefont {G.}~\bibnamefont {Moser}},\
  }\href {\doibase 10.1103/PhysRevLett.75.2706} {\bibfield  {journal} {\bibinfo
   {journal} {Phys. Rev. Lett.}\ }\textbf {\bibinfo {volume} {75}},\ \bibinfo
  {pages} {2706} (\bibinfo {year} {1995}{\natexlab{b}})}\BibitemShut {NoStop}%
\bibitem [{\citenamefont {Filippov}(1968)}]{filippov1968}%
  \BibitemOpen
  \bibfield  {author} {\bibinfo {author} {\bibfnamefont {L.~P.}\ \bibnamefont
  {Filippov}},\ }\href
  {http://www.sciencedirect.com/science/article/pii/0017931068901610}
  {\bibfield  {journal} {\bibinfo  {journal} {Int. J. Heat Mass Transfer}\
  }\textbf {\bibinfo {volume} {11}},\ \bibinfo {pages} {331} (\bibinfo {year}
  {1968})}\BibitemShut {NoStop}%
\bibitem [{\citenamefont {Hao}\ \emph {et~al.}(2005)\citenamefont {Hao},
  \citenamefont {Ferrell},\ and\ \citenamefont {Bhattacharjee}}]{hao2005}%
  \BibitemOpen
  \bibfield  {author} {\bibinfo {author} {\bibfnamefont {H.}~\bibnamefont
  {Hao}}, \bibinfo {author} {\bibfnamefont {R.~A.}\ \bibnamefont {Ferrell}}, \
  and\ \bibinfo {author} {\bibfnamefont {J.~K.}\ \bibnamefont
  {Bhattacharjee}},\ }\href {\doibase 10.1103/PhysRevE.71.021201} {\bibfield
  {journal} {\bibinfo  {journal} {Phys. Rev. E}\ }\textbf {\bibinfo {volume}
  {71}},\ \bibinfo {pages} {021201} (\bibinfo {year} {2005})}\BibitemShut
  {NoStop}%
\bibitem [{\citenamefont {Bhattacharjee}\ and\ \citenamefont
  {Ferrell}(1982)}]{bhattacharjee1982}%
  \BibitemOpen
  \bibfield  {author} {\bibinfo {author} {\bibfnamefont {J.~K.}\ \bibnamefont
  {Bhattacharjee}}\ and\ \bibinfo {author} {\bibfnamefont {R.~A.}\ \bibnamefont
  {Ferrell}},\ }\href {\doibase 10.1016/0375-9601(82)90595-3} {\bibfield
  {journal} {\bibinfo  {journal} {Phys. Lett. A}\ }\textbf {\bibinfo {volume}
  {88}},\ \bibinfo {pages} {77 } (\bibinfo {year} {1982})}\BibitemShut
  {NoStop}%
\bibitem [{\citenamefont {Paladin}\ and\ \citenamefont
  {Peliti}(1982)}]{paladin1982}%
  \BibitemOpen
  \bibfield  {author} {\bibinfo {author} {\bibfnamefont {G.}~\bibnamefont
  {Paladin}}\ and\ \bibinfo {author} {\bibfnamefont {L.}~\bibnamefont
  {Peliti}},\ }\href {\doibase 10.1051/jphyslet:0198200430101500} {\bibfield
  {journal} {\bibinfo  {journal} {J. Physique Lett.}\ }\textbf {\bibinfo
  {volume} {43}},\ \bibinfo {pages} {15} (\bibinfo {year} {1982})}\BibitemShut
  {NoStop}%
\bibitem [{\citenamefont {Halperin}\ \emph {et~al.}(1976)\citenamefont
  {Halperin}, \citenamefont {Hohenberg},\ and\ \citenamefont
  {Siggia}}]{halperin1976}%
  \BibitemOpen
  \bibfield  {author} {\bibinfo {author} {\bibfnamefont {B.~I.}\ \bibnamefont
  {Halperin}}, \bibinfo {author} {\bibfnamefont {P.~C.}\ \bibnamefont
  {Hohenberg}}, \ and\ \bibinfo {author} {\bibfnamefont {E.~D.}\ \bibnamefont
  {Siggia}},\ }\href {\doibase 10.1103/PhysRevB.13.1299} {\bibfield  {journal}
  {\bibinfo  {journal} {Phys. Rev. B}\ }\textbf {\bibinfo {volume} {13}},\
  \bibinfo {pages} {1299} (\bibinfo {year} {1976})}\BibitemShut {NoStop}%
\bibitem [{\citenamefont {Berg}\ \emph {et~al.}(1999)\citenamefont {Berg},
  \citenamefont {Moldover},\ and\ \citenamefont {Zimmerli}}]{berg1999}%
  \BibitemOpen
  \bibfield  {author} {\bibinfo {author} {\bibfnamefont {R.~F.}\ \bibnamefont
  {Berg}}, \bibinfo {author} {\bibfnamefont {M.~R.}\ \bibnamefont {Moldover}},
  \ and\ \bibinfo {author} {\bibfnamefont {G.~A.}\ \bibnamefont {Zimmerli}},\
  }\href {\doibase 10.1103/PhysRevE.60.4079} {\bibfield  {journal} {\bibinfo
  {journal} {Phys. Rev. E}\ }\textbf {\bibinfo {volume} {60}},\ \bibinfo
  {pages} {4079} (\bibinfo {year} {1999})}\BibitemShut {NoStop}%
\bibitem [{\citenamefont {Fisher}(1968)}]{fisher1968}%
  \BibitemOpen
  \bibfield  {author} {\bibinfo {author} {\bibfnamefont {M.~E.}\ \bibnamefont
  {Fisher}},\ }\href {\doibase 10.1103/PhysRev.176.257} {\bibfield  {journal}
  {\bibinfo  {journal} {Phys. Rev.}\ }\textbf {\bibinfo {volume} {176}},\
  \bibinfo {pages} {257} (\bibinfo {year} {1968})}\BibitemShut {NoStop}%
\bibitem [{\citenamefont {Ohta}\ and\ \citenamefont
  {Kawasaki}(1975)}]{ohta1975}%
  \BibitemOpen
  \bibfield  {author} {\bibinfo {author} {\bibfnamefont {T.}~\bibnamefont
  {Ohta}}\ and\ \bibinfo {author} {\bibfnamefont {K.}~\bibnamefont
  {Kawasaki}},\ }\href {\doibase 10.1143/PTP.55.1384} {\bibfield  {journal}
  {\bibinfo  {journal} {Prog. Theor. Phys.}\ }\textbf {\bibinfo {volume}
  {55}},\ \bibinfo {pages} {1384} (\bibinfo {year} {1975})}\BibitemShut
  {NoStop}%
\bibitem [{\citenamefont {Kadanoff}\ and\ \citenamefont
  {Swift}(1968)}]{kadanoff1968}%
  \BibitemOpen
  \bibfield  {author} {\bibinfo {author} {\bibfnamefont {L.~P.}\ \bibnamefont
  {Kadanoff}}\ and\ \bibinfo {author} {\bibfnamefont {J.}~\bibnamefont
  {Swift}},\ }\href {\doibase 10.1103/PhysRev.166.89} {\bibfield  {journal}
  {\bibinfo  {journal} {Phys. Rev.}\ }\textbf {\bibinfo {volume} {166}},\
  \bibinfo {pages} {89} (\bibinfo {year} {1968})}\BibitemShut {NoStop}%
\bibitem [{\citenamefont {Wilson}\ and\ \citenamefont
  {Fisher}(1972)}]{wilson1972}%
  \BibitemOpen
  \bibfield  {author} {\bibinfo {author} {\bibfnamefont {K.~G.}\ \bibnamefont
  {Wilson}}\ and\ \bibinfo {author} {\bibfnamefont {M.~E.}\ \bibnamefont
  {Fisher}},\ }\href {\doibase 10.1103/PhysRevLett.28.240} {\bibfield
  {journal} {\bibinfo  {journal} {Phys. Rev. Lett.}\ }\textbf {\bibinfo
  {volume} {28}},\ \bibinfo {pages} {240} (\bibinfo {year} {1972})}\BibitemShut
  {NoStop}%
\bibitem [{\citenamefont {Bausch}\ \emph {et~al.}(1976)\citenamefont {Bausch},
  \citenamefont {Janssen},\ and\ \citenamefont {Wagner}}]{bausch1976}%
  \BibitemOpen
  \bibfield  {author} {\bibinfo {author} {\bibfnamefont {R.}~\bibnamefont
  {Bausch}}, \bibinfo {author} {\bibfnamefont {H.~K.}\ \bibnamefont {Janssen}},
  \ and\ \bibinfo {author} {\bibfnamefont {H.}~\bibnamefont {Wagner}},\ }\href
  {\doibase 10.1007/BF01312880} {\bibfield  {journal} {\bibinfo  {journal} {Z.
  Phys. B}\ }\textbf {\bibinfo {volume} {24}},\ \bibinfo {pages} {113}
  (\bibinfo {year} {1976})}\BibitemShut {NoStop}%
\bibitem [{\citenamefont {Burstyn}\ \emph {et~al.}(1983)\citenamefont
  {Burstyn}, \citenamefont {Sengers}, \citenamefont {Bhattacharjee},\ and\
  \citenamefont {Ferrell}}]{burstyn1983}%
  \BibitemOpen
  \bibfield  {author} {\bibinfo {author} {\bibfnamefont {H.~C.}\ \bibnamefont
  {Burstyn}}, \bibinfo {author} {\bibfnamefont {J.~V.}\ \bibnamefont
  {Sengers}}, \bibinfo {author} {\bibfnamefont {J.~K.}\ \bibnamefont
  {Bhattacharjee}}, \ and\ \bibinfo {author} {\bibfnamefont {R.~A.}\
  \bibnamefont {Ferrell}},\ }\href {\doibase 10.1103/PhysRevA.28.1567}
  {\bibfield  {journal} {\bibinfo  {journal} {Phys. Rev. A}\ }\textbf {\bibinfo
  {volume} {28}},\ \bibinfo {pages} {1567} (\bibinfo {year}
  {1983})}\BibitemShut {NoStop}%
\bibitem [{\citenamefont {Swift}(1968)}]{swift1968}%
  \BibitemOpen
  \bibfield  {author} {\bibinfo {author} {\bibfnamefont {J.}~\bibnamefont
  {Swift}},\ }\href {\doibase 10.1103/PhysRev.173.257} {\bibfield  {journal}
  {\bibinfo  {journal} {Phys. Rev.}\ }\textbf {\bibinfo {volume} {173}},\
  \bibinfo {pages} {257} (\bibinfo {year} {1968})}\BibitemShut {NoStop}%
\bibitem [{\citenamefont {Gillis}\ \emph {et~al.}(2005)\citenamefont {Gillis},
  \citenamefont {Shinder},\ and\ \citenamefont {Moldover}}]{gillis2005}%
  \BibitemOpen
  \bibfield  {author} {\bibinfo {author} {\bibfnamefont {K.~A.}\ \bibnamefont
  {Gillis}}, \bibinfo {author} {\bibfnamefont {I.~I.}\ \bibnamefont {Shinder}},
  \ and\ \bibinfo {author} {\bibfnamefont {M.~R.}\ \bibnamefont {Moldover}},\
  }\href {\doibase 10.1103/PhysRevE.72.051201} {\bibfield  {journal} {\bibinfo
  {journal} {Phys. Rev. E}\ }\textbf {\bibinfo {volume} {72}},\ \bibinfo
  {pages} {051201} (\bibinfo {year} {2005})}\BibitemShut {NoStop}%
\bibitem [{\citenamefont {Jagannathan}\ and\ \citenamefont
  {Yethiraj}(2004)}]{jagannathan2004}%
  \BibitemOpen
  \bibfield  {author} {\bibinfo {author} {\bibfnamefont {K.}~\bibnamefont
  {Jagannathan}}\ and\ \bibinfo {author} {\bibfnamefont {A.}~\bibnamefont
  {Yethiraj}},\ }\href {\doibase 10.1103/PhysRevLett.93.015701} {\bibfield
  {journal} {\bibinfo  {journal} {Phys. Rev. Lett.}\ }\textbf {\bibinfo
  {volume} {93}},\ \bibinfo {pages} {015701} (\bibinfo {year}
  {2004})}\BibitemShut {NoStop}%
\bibitem [{\citenamefont {Chen}\ \emph {et~al.}(2005)\citenamefont {Chen},
  \citenamefont {Chimowitz}, \citenamefont {De},\ and\ \citenamefont
  {Shapir}}]{chen2005}%
  \BibitemOpen
  \bibfield  {author} {\bibinfo {author} {\bibfnamefont {A.}~\bibnamefont
  {Chen}}, \bibinfo {author} {\bibfnamefont {E.~H.}\ \bibnamefont {Chimowitz}},
  \bibinfo {author} {\bibfnamefont {S.}~\bibnamefont {De}}, \ and\ \bibinfo
  {author} {\bibfnamefont {Y.}~\bibnamefont {Shapir}},\ }\href {\doibase
  10.1103/PhysRevLett.95.255701} {\bibfield  {journal} {\bibinfo  {journal}
  {Phys. Rev. Lett.}\ }\textbf {\bibinfo {volume} {95}},\ \bibinfo {pages}
  {255701} (\bibinfo {year} {2005})}\BibitemShut {NoStop}%
\bibitem [{\citenamefont {Roy}\ and\ \citenamefont {Das}(2011)}]{roy2011}%
  \BibitemOpen
  \bibfield  {author} {\bibinfo {author} {\bibfnamefont {S.}~\bibnamefont
  {Roy}}\ and\ \bibinfo {author} {\bibfnamefont {S.~K.}\ \bibnamefont {Das}},\
  }\href {http://stacks.iop.org/0295-5075/94/i=3/a=36001} {\bibfield  {journal}
  {\bibinfo  {journal} {EPL (Europhys. Lett.)}\ }\textbf {\bibinfo {volume}
  {94}},\ \bibinfo {pages} {36001} (\bibinfo {year} {2011})}\BibitemShut
  {NoStop}%
\bibitem [{\citenamefont {Meier}\ \emph {et~al.}(2005)\citenamefont {Meier},
  \citenamefont {Laesecke},\ and\ \citenamefont {Kabelac}}]{meier2005}%
  \BibitemOpen
  \bibfield  {author} {\bibinfo {author} {\bibfnamefont {K.}~\bibnamefont
  {Meier}}, \bibinfo {author} {\bibfnamefont {A.}~\bibnamefont {Laesecke}}, \
  and\ \bibinfo {author} {\bibfnamefont {S.}~\bibnamefont {Kabelac}},\ }\href
  {\doibase 10.1063/1.1828040} {\bibfield  {journal} {\bibinfo  {journal} {J.
  Chem. Phys.}\ }\textbf {\bibinfo {volume} {122}},\ \bibinfo {eid} {014513}
  (\bibinfo {year} {2005})}\BibitemShut {NoStop}%
\bibitem [{\citenamefont {Dyer}\ \emph {et~al.}(2007)\citenamefont {Dyer},
  \citenamefont {Pettitt},\ and\ \citenamefont {Stell}}]{dyer2007}%
  \BibitemOpen
  \bibfield  {author} {\bibinfo {author} {\bibfnamefont {K.~M.}\ \bibnamefont
  {Dyer}}, \bibinfo {author} {\bibfnamefont {B.~M.}\ \bibnamefont {Pettitt}}, \
  and\ \bibinfo {author} {\bibfnamefont {G.}~\bibnamefont {Stell}},\ }\href
  {\doibase 10.1063/1.2424714} {\bibfield  {journal} {\bibinfo  {journal} {J.
  Chem. Phys.}\ }\textbf {\bibinfo {volume} {126}},\ \bibinfo {eid} {034502}
  (\bibinfo {year} {2007})}\BibitemShut {NoStop}%
\bibitem [{\citenamefont {Das}\ \emph {et~al.}(2006{\natexlab{a}})\citenamefont
  {Das}, \citenamefont {Fisher}, \citenamefont {Sengers}, \citenamefont
  {Horbach},\ and\ \citenamefont {Binder}}]{das2006prl}%
  \BibitemOpen
  \bibfield  {author} {\bibinfo {author} {\bibfnamefont {S.~K.}\ \bibnamefont
  {Das}}, \bibinfo {author} {\bibfnamefont {M.~E.}\ \bibnamefont {Fisher}},
  \bibinfo {author} {\bibfnamefont {J.~V.}\ \bibnamefont {Sengers}}, \bibinfo
  {author} {\bibfnamefont {J.}~\bibnamefont {Horbach}}, \ and\ \bibinfo
  {author} {\bibfnamefont {K.}~\bibnamefont {Binder}},\ }\href {\doibase
  10.1103/PhysRevLett.97.025702} {\bibfield  {journal} {\bibinfo  {journal}
  {Phys. Rev. Lett.}\ }\textbf {\bibinfo {volume} {97}},\ \bibinfo {pages}
  {025702} (\bibinfo {year} {2006}{\natexlab{a}})}\BibitemShut {NoStop}%
\bibitem [{\citenamefont {Das}\ \emph {et~al.}(2006{\natexlab{b}})\citenamefont
  {Das}, \citenamefont {Horbach}, \citenamefont {Binder}, \citenamefont
  {Fisher},\ and\ \citenamefont {Sengers}}]{das2006jcp}%
  \BibitemOpen
  \bibfield  {author} {\bibinfo {author} {\bibfnamefont {S.~K.}\ \bibnamefont
  {Das}}, \bibinfo {author} {\bibfnamefont {J.}~\bibnamefont {Horbach}},
  \bibinfo {author} {\bibfnamefont {K.}~\bibnamefont {Binder}}, \bibinfo
  {author} {\bibfnamefont {M.~E.}\ \bibnamefont {Fisher}}, \ and\ \bibinfo
  {author} {\bibfnamefont {J.~V.}\ \bibnamefont {Sengers}},\ }\href {\doibase
  10.1063/1.2215613} {\bibfield  {journal} {\bibinfo  {journal} {J. Chem.
  Phys.}\ }\textbf {\bibinfo {volume} {125}},\ \bibinfo {eid} {024506}
  (\bibinfo {year} {2006}{\natexlab{b}})}\BibitemShut {NoStop}%
\bibitem [{\citenamefont {Das}\ \emph {et~al.}(2007)\citenamefont {Das},
  \citenamefont {Sengers},\ and\ \citenamefont {Fisher}}]{das2007jcp}%
  \BibitemOpen
  \bibfield  {author} {\bibinfo {author} {\bibfnamefont {S.~K.}\ \bibnamefont
  {Das}}, \bibinfo {author} {\bibfnamefont {J.~V.}\ \bibnamefont {Sengers}}, \
  and\ \bibinfo {author} {\bibfnamefont {M.~E.}\ \bibnamefont {Fisher}},\
  }\href
  {http://scitation.aip.org/content/aip/journal/jcp/127/14/10.1063/1.2770736}
  {\bibfield  {journal} {\bibinfo  {journal} {J. Chem. Phys.}\ }\textbf
  {\bibinfo {volume} {127}},\ \bibinfo {eid} {144506} (\bibinfo {year}
  {2007})}\BibitemShut {NoStop}%
\bibitem [{\citenamefont {Fisher}(1971)}]{fisher1971}%
  \BibitemOpen
  \bibfield  {author} {\bibinfo {author} {\bibfnamefont {M.}~\bibnamefont
  {Fisher}},\ }in\ \href@noop {} {\emph {\bibinfo {booktitle} {Critical
  Phenomena}}},\ \bibinfo {series} {Proc. Enrico Fermi Int. School of Physics},
  Vol.~\bibinfo {volume} {51},\ \bibinfo {editor} {edited by\ \bibinfo {editor}
  {\bibfnamefont {M.}~\bibnamefont {Green}}}\ (\bibinfo  {publisher}
  {Academic},\ \bibinfo {address} {New York},\ \bibinfo {year} {1971})\ pp.\
  \bibinfo {pages} {1--99}\BibitemShut {NoStop}%
\bibitem [{\citenamefont {Roy}\ and\ \citenamefont {Das}(2013)}]{roy2013}%
  \BibitemOpen
  \bibfield  {author} {\bibinfo {author} {\bibfnamefont {S.}~\bibnamefont
  {Roy}}\ and\ \bibinfo {author} {\bibfnamefont {S.~K.}\ \bibnamefont {Das}},\
  }\href {\doibase 10.1063/1.4817777} {\bibfield  {journal} {\bibinfo
  {journal} {J. Chem. Phys.}\ }\textbf {\bibinfo {volume} {139}},\ \bibinfo
  {eid} {064505} (\bibinfo {year} {2013})}\BibitemShut {NoStop}%
\bibitem [{\citenamefont {Frenkel}\ and\ \citenamefont
  {Smit}(2002)}]{frenkel2002}%
  \BibitemOpen
  \bibfield  {author} {\bibinfo {author} {\bibfnamefont {D.}~\bibnamefont
  {Frenkel}}\ and\ \bibinfo {author} {\bibfnamefont {B.}~\bibnamefont {Smit}},\
  }\href@noop {} {\emph {\bibinfo {title} {Understanding Molecular Simulations:
  From Algorithm to Applications}}}\ (\bibinfo  {publisher} {Academic},\
  \bibinfo {address} {San Diego},\ \bibinfo {year} {2002})\BibitemShut
  {NoStop}%
\bibitem [{\citenamefont {Roy}\ and\ \citenamefont {Das}(2014)}]{roy2014}%
  \BibitemOpen
  \bibfield  {author} {\bibinfo {author} {\bibfnamefont {S.}~\bibnamefont
  {Roy}}\ and\ \bibinfo {author} {\bibfnamefont {S.~K.}\ \bibnamefont {Das}},\
  }\href {\doibase 10.1063/1.4903810} {\bibfield  {journal} {\bibinfo
  {journal} {J. Chem. Phys.}\ }\textbf {\bibinfo {volume} {141}},\ \bibinfo
  {eid} {234502} (\bibinfo {year} {2014})}\BibitemShut {NoStop}%
\bibitem [{\citenamefont {Voigtmann}\ and\ \citenamefont
  {Horbach}(2009)}]{voigtmann2009}%
  \BibitemOpen
  \bibfield  {author} {\bibinfo {author} {\bibfnamefont {{\relax
  T}.}~\bibnamefont {Voigtmann}}\ and\ \bibinfo {author} {\bibfnamefont
  {J.}~\bibnamefont {Horbach}},\ }\href {\doibase
  10.1103/PhysRevLett.103.205901} {\bibfield  {journal} {\bibinfo  {journal}
  {Phys. Rev. Lett.}\ }\textbf {\bibinfo {volume} {103}},\ \bibinfo {eid}
  {205901} (\bibinfo {year} {2009})}\BibitemShut {NoStop}%
\bibitem [{\citenamefont {Zausch}\ \emph {et~al.}(2010)\citenamefont {Zausch},
  \citenamefont {Horbach}, \citenamefont {Virnau},\ and\ \citenamefont
  {Binder}}]{zausch2010}%
  \BibitemOpen
  \bibfield  {author} {\bibinfo {author} {\bibfnamefont {J.}~\bibnamefont
  {Zausch}}, \bibinfo {author} {\bibfnamefont {J.}~\bibnamefont {Horbach}},
  \bibinfo {author} {\bibfnamefont {P.}~\bibnamefont {Virnau}}, \ and\ \bibinfo
  {author} {\bibfnamefont {K.}~\bibnamefont {Binder}},\ }\href
  {http://stacks.iop.org/0953-8984/22/i=10/a=104120} {\bibfield  {journal}
  {\bibinfo  {journal} {J. Phys.: Condens. Matter}\ }\textbf {\bibinfo {volume}
  {22}},\ \bibinfo {pages} {104120} (\bibinfo {year} {2010})}\BibitemShut
  {NoStop}%
\bibitem [{\citenamefont {Colberg}\ and\ \citenamefont
  {H\"ofling}(2011)}]{colberg2011}%
  \BibitemOpen
  \bibfield  {author} {\bibinfo {author} {\bibfnamefont {P.~H.}\ \bibnamefont
  {Colberg}}\ and\ \bibinfo {author} {\bibfnamefont {F.}~\bibnamefont
  {H\"ofling}},\ }\href {\doibase 10.1016/j.cpc.2011.01.009} {\bibfield
  {journal} {\bibinfo  {journal} {Comput. Phys. Commun.}\ }\textbf {\bibinfo
  {volume} {182}},\ \bibinfo {pages} {1120} (\bibinfo {year}
  {2011})}\BibitemShut {NoStop}%
\bibitem [{\citenamefont {Weeks}\ \emph {et~al.}(1971)\citenamefont {Weeks},
  \citenamefont {Chandler},\ and\ \citenamefont {Andersen}}]{weeks1971}%
  \BibitemOpen
  \bibfield  {author} {\bibinfo {author} {\bibfnamefont {J.~D.}\ \bibnamefont
  {Weeks}}, \bibinfo {author} {\bibfnamefont {D.}~\bibnamefont {Chandler}}, \
  and\ \bibinfo {author} {\bibfnamefont {H.~C.}\ \bibnamefont {Andersen}},\
  }\href {\doibase 10.1063/1.1674820} {\bibfield  {journal} {\bibinfo
  {journal} {J. Chem. Phys.}\ }\textbf {\bibinfo {volume} {54}},\ \bibinfo
  {pages} {5237} (\bibinfo {year} {1971})}\BibitemShut {NoStop}%
\bibitem [{\citenamefont {Widom}\ and\ \citenamefont
  {Rowlinson}(1970)}]{widom1970}%
  \BibitemOpen
  \bibfield  {author} {\bibinfo {author} {\bibfnamefont {B.}~\bibnamefont
  {Widom}}\ and\ \bibinfo {author} {\bibfnamefont {J.~S.}\ \bibnamefont
  {Rowlinson}},\ }\href {\doibase 10.1063/1.1673203} {\bibfield  {journal}
  {\bibinfo  {journal} {J. Chem. Phys.}\ }\textbf {\bibinfo {volume} {52}},\
  \bibinfo {pages} {1670} (\bibinfo {year} {1970})}\BibitemShut {NoStop}%
\bibitem [{\citenamefont {T{\"a}uber}(2014)}]{tauber2014}%
  \BibitemOpen
  \bibfield  {author} {\bibinfo {author} {\bibfnamefont {U.}~\bibnamefont
  {T{\"a}uber}},\ }\href@noop {} {\emph {\bibinfo {title} {Critical Dynamics: A
  Field Theory Approach to Equilibrium and Non-Equilibrium Scaling Behavior}}}\
  (\bibinfo  {publisher} {Cambridge University Press},\ \bibinfo {address}
  {London},\ \bibinfo {year} {2014})\BibitemShut {NoStop}%
\bibitem [{\citenamefont {Allen}\ and\ \citenamefont
  {Tildesley}(1987)}]{allen1987}%
  \BibitemOpen
  \bibfield  {author} {\bibinfo {author} {\bibfnamefont {M.}~\bibnamefont
  {Allen}}\ and\ \bibinfo {author} {\bibfnamefont {D.}~\bibnamefont
  {Tildesley}},\ }\href@noop {} {\emph {\bibinfo {title} {Computer Simulations
  of Liquids}}}\ (\bibinfo  {publisher} {Clarendon},\ \bibinfo {address}
  {Oxford},\ \bibinfo {year} {1987})\BibitemShut {NoStop}%
\bibitem [{\citenamefont {Baker}\ and\ \citenamefont
  {Hirst}(2011)}]{baker2011}%
  \BibitemOpen
  \bibfield  {author} {\bibinfo {author} {\bibfnamefont {J.~A.}\ \bibnamefont
  {Baker}}\ and\ \bibinfo {author} {\bibfnamefont {J.~D.}\ \bibnamefont
  {Hirst}},\ }\href {\doibase 10.1002/minf.201100042} {\bibfield  {journal}
  {\bibinfo  {journal} {Mol. Inform.}\ }\textbf {\bibinfo {volume} {30}},\
  \bibinfo {pages} {498} (\bibinfo {year} {2011})}\BibitemShut {NoStop}%
\bibitem [{\citenamefont {Weigel}\ \emph {et~al.}(2012)\citenamefont {Weigel},
  \citenamefont {Arnold},\ and\ \citenamefont {Virnau}}]{weigel2012}%
  \BibitemOpen
  \bibfield  {author} {\bibinfo {author} {\bibfnamefont {M.}~\bibnamefont
  {Weigel}}, \bibinfo {author} {\bibfnamefont {A.}~\bibnamefont {Arnold}}, \
  and\ \bibinfo {author} {\bibfnamefont {P.}~\bibnamefont {Virnau}},\ }\href
  {\doibase 10.1140/epjst/e2012-01633-0} {\bibfield  {journal} {\bibinfo
  {journal} {Eur. Phys. J. Special Topics}\ }\textbf {\bibinfo {volume}
  {210}},\ \bibinfo {pages} {1} (\bibinfo {year} {2012})}\BibitemShut {NoStop}%
\bibitem [{HAL(2016)}]{HALMD}%
  \BibitemOpen
  \href@noop {} {\enquote {\bibinfo {title} {{H}ighly {A}ccelerated
  {L}arge-scale {M}olecular {D}ynamics package},}\ } (\bibinfo {year}
  {2007--2016}),\ \bibinfo {note} {version 1.0, see
  \url{http://halmd.org}}\BibitemShut {NoStop}%
\bibitem [{\citenamefont {Ruymgaart}\ \emph {et~al.}(2011)\citenamefont
  {Ruymgaart}, \citenamefont {Cardenas},\ and\ \citenamefont
  {Elber}}]{ruymgaart2011}%
  \BibitemOpen
  \bibfield  {author} {\bibinfo {author} {\bibfnamefont {A.~P.}\ \bibnamefont
  {Ruymgaart}}, \bibinfo {author} {\bibfnamefont {A.~E.}\ \bibnamefont
  {Cardenas}}, \ and\ \bibinfo {author} {\bibfnamefont {R.}~\bibnamefont
  {Elber}},\ }\href {\doibase 10.1021/ct200360f} {\bibfield  {journal}
  {\bibinfo  {journal} {J. Chem. Theory Comput.}\ }\textbf {\bibinfo {volume}
  {7}},\ \bibinfo {pages} {3072} (\bibinfo {year} {2011})}\BibitemShut
  {NoStop}%
\bibitem [{\citenamefont {de~Buyl}\ \emph {et~al.}(2014)\citenamefont
  {de~Buyl}, \citenamefont {Colberg},\ and\ \citenamefont
  {H\"of{}ling}}]{debuyl2014}%
  \BibitemOpen
  \bibfield  {author} {\bibinfo {author} {\bibfnamefont {P.}~\bibnamefont
  {de~Buyl}}, \bibinfo {author} {\bibfnamefont {P.}~\bibnamefont {Colberg}}, \
  and\ \bibinfo {author} {\bibfnamefont {F.}~\bibnamefont {H\"of{}ling}},\
  }\href {\doibase 10.1016/j.cpc.2014.01.018} {\bibfield  {journal} {\bibinfo
  {journal} {Comput. Phys. Commun.}\ }\textbf {\bibinfo {volume} {185}},\
  \bibinfo {pages} {1546} (\bibinfo {year} {2014})}\BibitemShut {NoStop}%
\bibitem [{\citenamefont {Höf{}ling}\ and\ \citenamefont
  {Dietrich}(2015)}]{hoefling2015}%
  \BibitemOpen
  \bibfield  {author} {\bibinfo {author} {\bibfnamefont {F.}~\bibnamefont
  {Höf{}ling}}\ and\ \bibinfo {author} {\bibfnamefont {S.}~\bibnamefont
  {Dietrich}},\ }\href {\doibase 10.1209/0295-5075/109/46002} {\bibfield
  {journal} {\bibinfo  {journal} {EPL (Europhys. Lett.)}\ }\textbf {\bibinfo
  {volume} {109}},\ \bibinfo {pages} {46002} (\bibinfo {year}
  {2015})}\BibitemShut {NoStop}%
\bibitem [{\citenamefont {Martyna}\ \emph {et~al.}(1992)\citenamefont
  {Martyna}, \citenamefont {Klein},\ and\ \citenamefont
  {Tuckerman}}]{martyna1992}%
  \BibitemOpen
  \bibfield  {author} {\bibinfo {author} {\bibfnamefont {G.~J.}\ \bibnamefont
  {Martyna}}, \bibinfo {author} {\bibfnamefont {M.~L.}\ \bibnamefont {Klein}},
  \ and\ \bibinfo {author} {\bibfnamefont {M.}~\bibnamefont {Tuckerman}},\
  }\href {\doibase 10.1063/1.463940} {\bibfield  {journal} {\bibinfo  {journal}
  {J. Chem. Phys.}\ }\textbf {\bibinfo {volume} {97}},\ \bibinfo {pages} {2635}
  (\bibinfo {year} {1992})}\BibitemShut {NoStop}%
\bibitem [{\citenamefont {Das}\ \emph {et~al.}(2003)\citenamefont {Das},
  \citenamefont {Horbach},\ and\ \citenamefont {Binder}}]{das2003}%
  \BibitemOpen
  \bibfield  {author} {\bibinfo {author} {\bibfnamefont {S.~K.}\ \bibnamefont
  {Das}}, \bibinfo {author} {\bibfnamefont {J.}~\bibnamefont {Horbach}}, \ and\
  \bibinfo {author} {\bibfnamefont {K.}~\bibnamefont {Binder}},\ }\href
  {\doibase 10.1063/1.1580106} {\bibfield  {journal} {\bibinfo  {journal} {J.
  Chem. Phys.}\ }\textbf {\bibinfo {volume} {119}},\ \bibinfo {pages} {1547}
  (\bibinfo {year} {2003})}\BibitemShut {NoStop}%
\bibitem [{\citenamefont {Toxvaerd}\ and\ \citenamefont
  {Velasco}(1995)}]{toxvaerd1995}%
  \BibitemOpen
  \bibfield  {author} {\bibinfo {author} {\bibfnamefont {S.}~\bibnamefont
  {Toxvaerd}}\ and\ \bibinfo {author} {\bibfnamefont {E.}~\bibnamefont
  {Velasco}},\ }\href@noop {} {\bibfield  {journal} {\bibinfo  {journal} {Mol.
  Phys.}\ }\textbf {\bibinfo {volume} {86}},\ \bibinfo {pages} {845} (\bibinfo
  {year} {1995})}\BibitemShut {NoStop}%
\bibitem [{\citenamefont {Wilding}(1997)}]{wilding1997}%
  \BibitemOpen
  \bibfield  {author} {\bibinfo {author} {\bibfnamefont {N.~B.}\ \bibnamefont
  {Wilding}},\ }\href {\doibase 10.1103/PhysRevE.55.6624} {\bibfield  {journal}
  {\bibinfo  {journal} {Phys. Rev. E}\ }\textbf {\bibinfo {volume} {55}},\
  \bibinfo {pages} {6624} (\bibinfo {year} {1997})}\BibitemShut {NoStop}%
\bibitem [{\citenamefont {Dietrich}\ and\ \citenamefont
  {Latz}(1989)}]{dietrich1989}%
  \BibitemOpen
  \bibfield  {author} {\bibinfo {author} {\bibfnamefont {S.}~\bibnamefont
  {Dietrich}}\ and\ \bibinfo {author} {\bibfnamefont {A.}~\bibnamefont
  {Latz}},\ }\href {\doibase 10.1103/PhysRevB.40.9204} {\bibfield  {journal}
  {\bibinfo  {journal} {Phys. Rev. B}\ }\textbf {\bibinfo {volume} {40}},\
  \bibinfo {pages} {9204} (\bibinfo {year} {1989})}\BibitemShut {NoStop}%
\bibitem [{\citenamefont {Getta}\ and\ \citenamefont
  {Dietrich}(1993)}]{getta1993}%
  \BibitemOpen
  \bibfield  {author} {\bibinfo {author} {\bibfnamefont {T.}~\bibnamefont
  {Getta}}\ and\ \bibinfo {author} {\bibfnamefont {S.}~\bibnamefont
  {Dietrich}},\ }\href {http://link.aps.org/doi/10.1103/PhysRevE.47.1856}
  {\bibfield  {journal} {\bibinfo  {journal} {Phys. Rev. E}\ }\textbf {\bibinfo
  {volume} {47}},\ \bibinfo {pages} {1856} (\bibinfo {year}
  {1993})}\BibitemShut {NoStop}%
\bibitem [{\citenamefont {Dietrich}\ and\ \citenamefont
  {Schick}(1997)}]{dietrich1997}%
  \BibitemOpen
  \bibfield  {author} {\bibinfo {author} {\bibfnamefont {S.}~\bibnamefont
  {Dietrich}}\ and\ \bibinfo {author} {\bibfnamefont {M.}~\bibnamefont
  {Schick}},\ }\href
  {http://www.sciencedirect.com/science/article/pii/S0039602897001222}
  {\bibfield  {journal} {\bibinfo  {journal} {Surf. Sci.}\ }\textbf {\bibinfo
  {volume} {382}},\ \bibinfo {pages} {178} (\bibinfo {year}
  {1997})}\BibitemShut {NoStop}%
\bibitem [{\citenamefont {Hansen}\ and\ \citenamefont
  {McDonald}(2008)}]{hansen2008}%
  \BibitemOpen
  \bibfield  {author} {\bibinfo {author} {\bibfnamefont {J.-P.}\ \bibnamefont
  {Hansen}}\ and\ \bibinfo {author} {\bibfnamefont {I.}~\bibnamefont
  {McDonald}},\ }\href@noop {} {\emph {\bibinfo {title} {Theory of Simple
  Liquids}}}\ (\bibinfo  {publisher} {Academic},\ \bibinfo {address} {London},\
  \bibinfo {year} {2008})\BibitemShut {NoStop}%
\bibitem [{\citenamefont {Bhatia}\ and\ \citenamefont
  {Thronton}(1970)}]{bhatia1970}%
  \BibitemOpen
  \bibfield  {author} {\bibinfo {author} {\bibfnamefont {A.~B.}\ \bibnamefont
  {Bhatia}}\ and\ \bibinfo {author} {\bibfnamefont {D.~E.}\ \bibnamefont
  {Thronton}},\ }\href {\doibase 10.1103/PhysRevB.2.3004} {\bibfield  {journal}
  {\bibinfo  {journal} {Phys. Rev. B}\ }\textbf {\bibinfo {volume} {2}},\
  \bibinfo {pages} {8} (\bibinfo {year} {1970})}\BibitemShut {NoStop}%
\bibitem [{\citenamefont {Zhou}\ and\ \citenamefont {Miller}(1996)}]{zhou1996}%
  \BibitemOpen
  \bibfield  {author} {\bibinfo {author} {\bibfnamefont {Y.}~\bibnamefont
  {Zhou}}\ and\ \bibinfo {author} {\bibfnamefont {G.~H.}\ \bibnamefont
  {Miller}},\ }\href {\doibase 10.1021/jp9533739} {\bibfield  {journal}
  {\bibinfo  {journal} {J. Phys. Chem.}\ }\textbf {\bibinfo {volume} {100}},\
  \bibinfo {pages} {5516} (\bibinfo {year} {1996})}\BibitemShut {NoStop}%
\bibitem [{\citenamefont {Horbach}\ \emph {et~al.}(2007)\citenamefont
  {Horbach}, \citenamefont {Das}, \citenamefont {Griesche}, \citenamefont
  {Macht}, \citenamefont {Frohberg},\ and\ \citenamefont
  {Meyer}}]{horbach2007}%
  \BibitemOpen
  \bibfield  {author} {\bibinfo {author} {\bibfnamefont {J.}~\bibnamefont
  {Horbach}}, \bibinfo {author} {\bibfnamefont {S.~K.}\ \bibnamefont {Das}},
  \bibinfo {author} {\bibfnamefont {A.}~\bibnamefont {Griesche}}, \bibinfo
  {author} {\bibfnamefont {M.-P.}\ \bibnamefont {Macht}}, \bibinfo {author}
  {\bibfnamefont {G.}~\bibnamefont {Frohberg}}, \ and\ \bibinfo {author}
  {\bibfnamefont {A.}~\bibnamefont {Meyer}},\ }\href {\doibase
  10.1103/PhysRevB.75.174304} {\bibfield  {journal} {\bibinfo  {journal} {Phys.
  Rev. B}\ }\textbf {\bibinfo {volume} {75}},\ \bibinfo {pages} {174304}
  (\bibinfo {year} {2007})}\BibitemShut {NoStop}%
\bibitem [{\citenamefont {H{\"o}ft}(2012)}]{hoeft2012}%
  \BibitemOpen
  \bibfield  {author} {\bibinfo {author} {\bibfnamefont {N.}~\bibnamefont
  {H{\"o}ft}},\ }\href@noop {} {\emph {\bibinfo {title} {Diffusion dynamics in
  two-dimensional fluids}}},\ Master Thesis, Universität Düsseldorf, Germany\
  (\bibinfo {year} {2012})\BibitemShut {NoStop}%
\bibitem [{\citenamefont {H{\"o}f{}ling}\ and\ \citenamefont
  {Franosch}(2007)}]{hoefling2007}%
  \BibitemOpen
  \bibfield  {author} {\bibinfo {author} {\bibfnamefont {F.}~\bibnamefont
  {H{\"o}f{}ling}}\ and\ \bibinfo {author} {\bibfnamefont {T.}~\bibnamefont
  {Franosch}},\ }\href {\doibase 10.1103/PhysRevLett.98.140601} {\bibfield
  {journal} {\bibinfo  {journal} {Phys. Rev. Lett.}\ }\textbf {\bibinfo
  {volume} {98}},\ \bibinfo {eid} {140601} (\bibinfo {year}
  {2007})}\BibitemShut {NoStop}%
\bibitem [{\citenamefont {Sengers}(1985)}]{sengers1985}%
  \BibitemOpen
  \bibfield  {author} {\bibinfo {author} {\bibfnamefont {J.}~\bibnamefont
  {Sengers}},\ }\href {\doibase 10.1007/BF00522145} {\bibfield  {journal}
  {\bibinfo  {journal} {Int. J. Thermophys.}\ }\textbf {\bibinfo {volume}
  {6}},\ \bibinfo {pages} {203} (\bibinfo {year} {1985})}\BibitemShut {NoStop}%
\bibitem [{\citenamefont {Kawasaki}\ and\ \citenamefont
  {Lo}(1972)}]{kawasaki1972}%
  \BibitemOpen
  \bibfield  {author} {\bibinfo {author} {\bibfnamefont {K.}~\bibnamefont
  {Kawasaki}}\ and\ \bibinfo {author} {\bibfnamefont {S.-M.}\ \bibnamefont
  {Lo}},\ }\href {\doibase 10.1103/PhysRevLett.29.48} {\bibfield  {journal}
  {\bibinfo  {journal} {Phys. Rev. Lett.}\ }\textbf {\bibinfo {volume} {29}},\
  \bibinfo {pages} {48} (\bibinfo {year} {1972})}\BibitemShut {NoStop}%
\bibitem [{\citenamefont {Alder}\ \emph {et~al.}(1970)\citenamefont {Alder},
  \citenamefont {Gass},\ and\ \citenamefont {Wainwright}}]{alder1970}%
  \BibitemOpen
  \bibfield  {author} {\bibinfo {author} {\bibfnamefont {B.~J.}\ \bibnamefont
  {Alder}}, \bibinfo {author} {\bibfnamefont {D.~M.}\ \bibnamefont {Gass}}, \
  and\ \bibinfo {author} {\bibfnamefont {T.~E.}\ \bibnamefont {Wainwright}},\
  }\href {\doibase 10.1063/1.1673845} {\bibfield  {journal} {\bibinfo
  {journal} {J. Chem. Phys.}\ }\textbf {\bibinfo {volume} {53}},\ \bibinfo
  {pages} {3813} (\bibinfo {year} {1970})}\BibitemShut {NoStop}%
\bibitem [{\citenamefont {Viscardy}\ and\ \citenamefont
  {Gaspard}(2003)}]{viscardy2003}%
  \BibitemOpen
  \bibfield  {author} {\bibinfo {author} {\bibfnamefont {S.}~\bibnamefont
  {Viscardy}}\ and\ \bibinfo {author} {\bibfnamefont {P.}~\bibnamefont
  {Gaspard}},\ }\href {\doibase 10.1103/PhysRevE.68.041204} {\bibfield
  {journal} {\bibinfo  {journal} {Phys. Rev. E}\ }\textbf {\bibinfo {volume}
  {68}},\ \bibinfo {pages} {041204} (\bibinfo {year} {2003})}\BibitemShut
  {NoStop}%
\bibitem [{\citenamefont {Ohta}(1977)}]{ohta1977}%
  \BibitemOpen
  \bibfield  {author} {\bibinfo {author} {\bibfnamefont {T.}~\bibnamefont
  {Ohta}},\ }\href {http://stacks.iop.org/0022-3719/10/i=6/a=010} {\bibfield
  {journal} {\bibinfo  {journal} {J. Phys. C}\ }\textbf {\bibinfo {volume}
  {10}},\ \bibinfo {pages} {791} (\bibinfo {year} {1977})}\BibitemShut
  {NoStop}%
\bibitem [{\citenamefont {Jacobs}(1986)}]{jacobs1986}%
  \BibitemOpen
  \bibfield  {author} {\bibinfo {author} {\bibfnamefont {D.~T.}\ \bibnamefont
  {Jacobs}},\ }\href {\doibase 10.1103/PhysRevA.33.2605} {\bibfield  {journal}
  {\bibinfo  {journal} {Phys. Rev. A}\ }\textbf {\bibinfo {volume} {33}},\
  \bibinfo {pages} {2605} (\bibinfo {year} {1986})}\BibitemShut {NoStop}%
\bibitem [{\citenamefont {Mirzaev}\ \emph {et~al.}(2006)\citenamefont
  {Mirzaev}, \citenamefont {Iwanowski}, \citenamefont {Zaitdinov},\ and\
  \citenamefont {Kaatze}}]{mirzaev2006}%
  \BibitemOpen
  \bibfield  {author} {\bibinfo {author} {\bibfnamefont {S.}~\bibnamefont
  {Mirzaev}}, \bibinfo {author} {\bibfnamefont {I.}~\bibnamefont {Iwanowski}},
  \bibinfo {author} {\bibfnamefont {M.}~\bibnamefont {Zaitdinov}}, \ and\
  \bibinfo {author} {\bibfnamefont {U.}~\bibnamefont {Kaatze}},\ }\href
  {\doibase 10.1016/j.cplett.2006.10.013} {\bibfield  {journal} {\bibinfo
  {journal} {Chem. Phys. Lett.}\ }\textbf {\bibinfo {volume} {431}},\ \bibinfo
  {pages} {308 } (\bibinfo {year} {2006})}\BibitemShut {NoStop}%
\bibitem [{\citenamefont {Das}\ and\ \citenamefont {Binder}(2011)}]{das2011}%
  \BibitemOpen
  \bibfield  {author} {\bibinfo {author} {\bibfnamefont {S.~K.}\ \bibnamefont
  {Das}}\ and\ \bibinfo {author} {\bibfnamefont {K.}~\bibnamefont {Binder}},\
  }\href {\doibase 10.1103/physrevlett.107.235702} {\bibfield  {journal}
  {\bibinfo  {journal} {Phys. Rev. Lett.}\ }\textbf {\bibinfo {volume} {107}},\
  \bibinfo {pages} {235702} (\bibinfo {year} {2011})}\BibitemShut {NoStop}%
\bibitem [{\citenamefont {Guevara-Carrion}\ \emph {et~al.}(2016)\citenamefont
  {Guevara-Carrion}, \citenamefont {Janzen}, \citenamefont {Munoz-Munoz},\ and\
  \citenamefont {Vrabec}}]{guevara2016}%
  \BibitemOpen
  \bibfield  {author} {\bibinfo {author} {\bibfnamefont {G.}~\bibnamefont
  {Guevara-Carrion}}, \bibinfo {author} {\bibfnamefont {T.}~\bibnamefont
  {Janzen}}, \bibinfo {author} {\bibfnamefont {Y.~M.}\ \bibnamefont
  {Munoz-Munoz}}, \ and\ \bibinfo {author} {\bibfnamefont {J.}~\bibnamefont
  {Vrabec}},\ }\href
  {http://scitation.aip.org/content/aip/journal/jcp/144/12/10.1063/1.4943395}
  {\bibfield  {journal} {\bibinfo  {journal} {J. Chem. Phys.}\ }\textbf
  {\bibinfo {volume} {144}},\ \bibinfo {eid} {124501} (\bibinfo {year}
  {2016})}\BibitemShut {NoStop}%
\end{thebibliography}%

\end{document}